\documentclass[aps, prx, reprint, showpacs, superscriptaddress]{revtex4-2}
\usepackage{graphicx}
\usepackage{amsmath,amsfonts,amssymb,bm,amsthm,bbm}
\usepackage{hyperref}%
\usepackage{tikz}
\usepackage{changes}

\newcommand{\rd}{\textrm{d}}
\newcommand{\re}{\textrm{e}}
\newcommand{\inp}[2]{\langle#1|#2\rangle}
\newcommand{\ket}[1]{|#1\rangle}

\newcommand{\ave}[1]{\langle #1 \rangle}
\makeatletter
\newcommand*{\rom}[1]{\expandafter\@slowromancap\romannumeral #1@}
\makeatother

\begin{document}
\title{Giant High-order Nonlinear and Nonreciprocal Electrical Transports Induced by Valley Flipping in Bernal Bilayer Graphene}

\author{Yuelin Shao}
\email{ylshao@ust.hk}
\affiliation{Department of Physics, The Hongkong University of Science and Technology,
Clear Water Bay, Kowloon 999077, Hong Kong, China}
\affiliation{Beijing National Laboratory for Condensed Matter Physics and Institute of Physics, Chinese Academy of Sciences,
Beijing 100190, China}
\affiliation{School of Physical Sciences, University of Chinese Academy of Sciences, Beijing 100049, China}
\author{Xi Dai}
\email{daix@ust.hk}
\affiliation{Department of Physics, The Hongkong University of Science and Technology,
Clear Water Bay, Kowloon 999077, Hong Kong, China}

\date{\today}

\begin{abstract}
  We investigate the electrical transport properties of the mini-valley polarized state proposed recently in slightly doped Bernal Bilayer Graphene (BLG) in large electric displacement fields.
  By minimizing the Hartree-Fock (HF) energy functional, we first confirm the appearance of mini-valley polarized phase.
  At the low carrier doping regime, the 1-pocket state will be stabilized where only one of the trigonal-wrapping-induced Fermi pockets near the atomic-valley center is filled.
  Then we study the electrical transport of the 1-pocket state by solving the Boltzmann equation.
  We find that the valley polarization could be easily flopped by an in-plane electrical field, which will lead to hysteresis loop in the direct current (DC) I-V curves.
  Such irreversible current responses in the DC limit will directly induce strong nonlinear and nonreciprocal alternating current (AC) responses, which has been already observed in the recent experiments on BLG.

\end{abstract}
\maketitle

\section{Introduction}

In the realm of two-dimensional materials, the valley degree of freedom is a critical factor that influences their unique magnetic, transport, and optical characteristics. 
This is in addition to the well-known degrees of freedom in condensed matter, such as spin, orbital, and momentum. 
In certain systems, like twisted bilayer graphene (TBG) and transition metal dichalcogenides (TMDs), a variety of valley polarization and hybridization states have been proposed\cite{liuQuantumValleyHall2019,kangStrongCouplingPhases2019,zhangCorrelatedInsulatingPhases2020,xieNatureCorrelatedInsulator2020,liuTheoriesCorrelatedInsulating2021,devakulMagicTwistedTransition2021} and observed\cite{caoCorrelatedInsulatorBehaviour2018, sharpeEmergentFerromagnetismThreequarters2019a,jiangChargeOrderBroken2019,choiElectronicCorrelationsTwisted2019a,caoTunableCorrelatedStates2020,serlinIntrinsicQuantizedAnomalous2020,chenTunableCorrelatedChern2020,polshynElectricalSwitchingMagnetic2020,tschirhartImagingOrbitalFerromagnetism2021,choiCorrelationdrivenTopologicalPhases2021,pierceUnconventionalSequenceCorrelated2021}. 
These states contribute to the emergence of exotic phenomena, including the quantum anomalous Hall effect, correlated insulators, and orbital magnetism. 
Recently, valley-polarized states have also been detected in non-twisted systems such as Rhombohedral trilayer and Bernal bilayer graphene when a displacement field is applied along the $z$-axis\cite{zhouHalfQuartermetalsRhombohedral2021a,delabarreraCascadeIsospinPhase2022,seilerQuantumCascadeCorrelated2022,zhouIsospinMagnetismSpinpolarized2022,hanSpontaneousIsospinPolarization2023}. 
In this paper, we focus on the Bernal bilayer graphene (BLG) whose atomic structure and Brillouin zone is shown in Fig. \ref{fig:blg_non}(a).
The displacement field first separates the two Dirac nodes in energy and subsequently leads to the formation of a hybridization gap due to inter-layer coupling\cite{mccannAsymmetryGapElectronic2006,mccannLandauLevelDegeneracyQuantum2006,oostingaGateinducedInsulatingState2008}, resulting in a ``Mexican hat'' type band dispersion for each atomic valley $\pm K$. 
The trigonal warping effect in BLG further creates three local maxima in the valence band (or minima in the conduction band) near each atomic valley center\cite{mccannLandauLevelDegeneracyQuantum2006}, as indicated by $Q_{1,2,3}$ in Figure \ref{fig:blg_non}(b). 
As a result, the highest valence band (HVB) of gated BLG hosts six band maxima $\pm Q_{1,2,3}$ (the prefactor $\xi=\pm$ label the atomic valley, the subscript $\alpha=1,2,3$ label the mini-valley around the atomic valleys), which from now on are called all together as the valleys thought out this paper for simplicity.
When the filling factor of the system is just slightly away from the charge neutrality point, quantum oscillation experiments indicate an emergence of the half- and quarter-metal phases\cite{delabarreraCascadeIsospinPhase2022,seilerQuantumCascadeCorrelated2022,zhouIsospinMagnetismSpinpolarized2022}.
Starting from the quarter-metal phase where only one spin and atomic valley is occupied, if the doping concentration is further reduced so that the mini-valley $Q_{1,2,3}$ could also be viewed as a degree of freedom, a polarized state to the mini-valley is further predicted\cite{dongIsospinMomentumpolarizedOrders2023}.
Such kind of valley-polarized states break the threefold rotation symmetry $C_3$ and has been identified by the $C_3$-breaking pattern of angle-resolved second-harmonic nonlinear transport measurements\cite{linSpontaneousMomentumPolarization2023}.
More recently, anomalous high-order nonlinear and nonreciprocal transport signals have also been detected in the slightly doped BLG\cite{zanObservationElectricalHighharmonic2024}, challenging traditional transport theories and spurring the need for theoretical investigation from new vantage points. 

In this paper, we first establish the presence of mini-valley-polarized phases in gated BLG. 
This is deduced from the minimization of the system's total energy functional under Hartree-Fock (HF) approximation. 
The proximity of the three valleys around the $K$ points makes the polarized states quite susceptible to the transitions induced by current flow, potentially leading to significant nonlinear and non-reciprocal transport behaviors. 
We then analyze the current response of systems exhibiting "Mexican hat" type band dispersion by employing the Boltzmann equation, incorporating the effects of long-range Coulomb interactions among carriers. 
Differentiating between relaxation processes within the same mini-valley and across different valleys allowed us to derive equations that effectively describe the dynamics within these systems. 
The resulting phenomena from our analysis include: 
(1) The mini-valley polarization can be reversed via electric current, causing first-order hysteresis loops in the direct current (DC) I-V curve, which was also pointed out in a recent study\citep{panigrahiSignaturesElectronicOrdering2024a}. 
(2) The emergence of irreversible current responses in the DC spectrum, resulting in significant nonlinear and nonreciprocal alternating current (AC) behaviors at low frequencies. 
(3) The presence of a critical field intensity threshold beyond which the system's long-term current response becomes insensitive to initial polarization directions, yielding nearly symmetric angular responses.

We stress that the first-order hysteretic behaviors in the DC I-V curves and the related phenomena discussed above persist as long as the ground state is mini-valley polarized.
This means that these phenomena are also ``smoking gun'' for confirming the mini-valley polarized ground state.

{Similar current-induced hysteretic phenomena were also observed\citep{serlinIntrinsicQuantizedAnomalous2020,sharpeEmergentFerromagnetismThreequarters2019a} and studied\citep{suCurrentInducedReversalAnomalous2020,heGiantOrbitalMagnetoelectric2020} in TBG.
Nonetheless, the mechanisms in TBG are quite different from this work.
Firstly, the ground state in TBG is an insulator while the ground state in this study is a metal.
Besides, as will be shown in Sec.\ref{sec:dc}, the valley flipping phenomenon in our work results from the competition between the electric field driving effect and the inter-valley relaxation.
However, neither of these two processes could possibly exist in any of the insulator phase, such as the previous theoretical studies in TBG\citep{suCurrentInducedReversalAnomalous2020,heGiantOrbitalMagnetoelectric2020}.
In other words, we present a new mechanism of current-induced valley polarization flipping phenomenon in the valley polarized metal.
}


\section{The Valley-Polarized Ground States}\label{sec:polarized_gs}

\begin{figure}
  \centering
  \includegraphics[width=\linewidth]{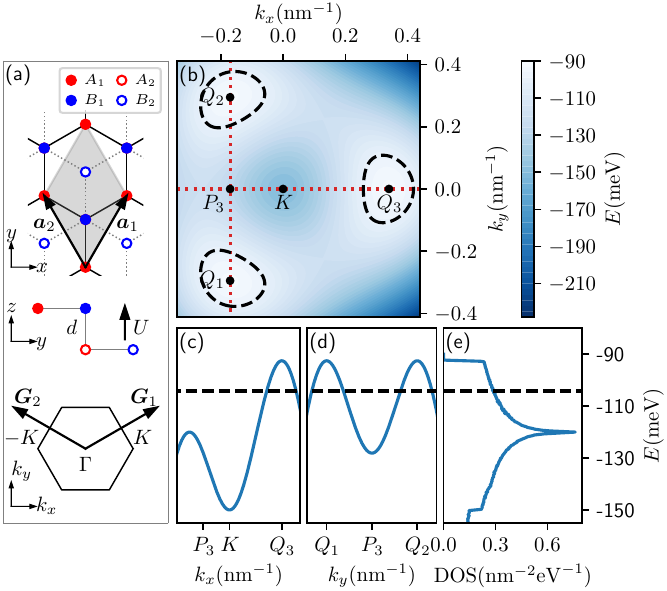}
  \caption{(a) Real space structure and Brillouin zone of BLG. 
  (b) Constant energy surface of the HVB in the $K$ valley for inter-layer bias potential $U=300\mathrm{meV}$.
  (c)(d) Detail view of the band structure along the tow dotted red lines in (b).
  (e) DOS per spin and per atomic valley.
  The dashed black lines in (b-e) mark the same energy level where the hole concentration contributed by each pocket in (b) is about $0.1\times 10^{12}\mathrm{cm}^{-2}$.
  }
  \label{fig:blg_non}
\end{figure}

The non-interacting band structures near the atomic valley center $\xi K$ ($\xi=\pm$) could be described by the following $k\cdot p$ Hamiltonian\cite{mccannElectronicPropertiesBilayer2013,jungAccurateTightbindingModels2014} under atomic basis $\{A_1,B_1,A_2,B_2\}$
\begin{equation}
  H_{\xi,\bm{k}}\approx \begin{bmatrix}
      -U/2              & v_0 \pi^{\dagger} & -v_4 \pi^{\dagger} & v_3 \pi    \\
      v_0 \pi           & -U/2 + \Delta'    & \gamma_1           & -v_4 \pi^{\dagger}\\
      -v_4\pi & \gamma_1          & U/2 +\Delta'       & v_0 \pi^{\dagger}  \\ 
      v_3 \pi^{\dagger} & -v_4 \pi          & v_0 \pi            & U/2                 
  \end{bmatrix}
\end{equation}
where $\pi=\hbar(\xi k_x+ik_y)$ and $U$ is the inter-layer potential arose from gating. 
The velocities $v_i$ are defined by the nearest neighbor hopping parameters $\gamma_i$ as $v_i=\sqrt{3}a\gamma_i/2\hbar$ and $a$ is the lattice constant (the hopping parameters in this paper is adopted from \citet{kuzmenkoDeterminationGatetunableBand2009}).
In this paper, we focus on the hole doping regime of the gated BLG and the HVB dominates the low energy excitations.
Taking $U=300\mathrm{meV}$, the constant energy contour of the HVB near $K$ point is plotted in Fig. \ref{fig:blg_non}(b) by the pseudo-color map.
And the band structures near the $-K$ valley are related with $K$ valley by time reversal symmetry.
The detail band structures along the two dotted red lines in Fig. \ref{fig:blg_non}(b) are shown in Fig. \ref{fig:blg_non}(c)(d).
In Fig. \ref{fig:blg_non}(e), we also plot the density of states (DOS) per-valley and per-spin and the Van Hove singularities are clearly shown.
Due to the trigonal-wrapping effect from the inter-layer coupling $\gamma_3$ between $A_1$ and $B_2$ atoms, the non-interacting Fermi surfaces for very low hole doping separate into three hole pockets as illustrated by the dashed black lines in Fig. \ref{fig:blg_non}(b).

In the low doping regime, the quarter-metal phase of gated BLG is verified both experimentally\cite{delabarreraCascadeIsospinPhase2022,seilerQuantumCascadeCorrelated2022} and theoretical\cite{xieFlavorSymmetryBreaking2023,dongIsospinMomentumpolarizedOrders2023,jangChiralityCorrelationsSpontaneous2023}.
We focus on the density region where the quarter-metal state is stable and assume the $(\uparrow,K)$ flavor is occupied, then the low energy physics could be described by the HVB electron in $(\uparrow,K)$ flavor with creation operator $c^{\dagger}_{\bm{k}}$.
By projecting to the occupied flavor, the many-body Hamiltonian could be approximately written as 
\begin{equation}
  H=\sum_{\bm{k}}\varepsilon_{\bm{k}}^0c^{\dagger}_{\bm{k}}c_{\bm{k}}+\frac{1}{2\mathcal{S}}\sum_{\bm{k}\bm{k}'\bm{q}}U_{\bm{k}\bm{k}'}(\bm{q})c^{\dagger}_{\bm{k}}c^{\dagger}_{\bm{k}'}c_{\bm{k}'+\bm{q}}c_{\bm{k}-\bm{q}},
\end{equation}
where $\mathcal{S}$ is the 2D area of the system.
The first term is the kinetic part and the non-interacting dispersion $\varepsilon^0_{\bm{k}}$ is shown in Fig. \ref{fig:blg_non}(b).
The second term is the interacting part, where $U_{\bm{k}\bm{k}'}(\bm{q})\equiv V(\bm{q})\inp{\bm{k}}{\bm{k}-\bm{q}}\inp{\bm{k}'}{\bm{k}'+\bm{q}}$ is the interaction matrix projected to HVB and $\ket{\bm{k}}$ is the wavefunction of HVB electron in $(\uparrow,K_{+})$ flavor.
To obtain the previous form of interaction, only the electron-electron scattering with momentum transfer much smaller than Brillouin zone, i.e. $q\ll G$, is considered, which is a reasonable approximation for long range interaction.
In this paper, the interaction $V(\bm{q})$ is taken as the Coulomb one $V_{col}(q)=2\pi/\epsilon q$ and the dielectric constant is set as $\epsilon=5$.

\begin{figure}
  \centering
  \includegraphics[width=\linewidth]{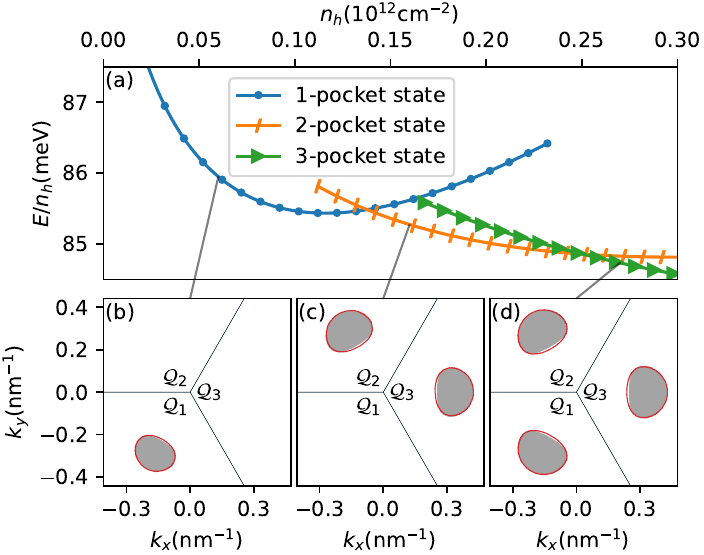}
  \caption{(a) Total energies per hole $E_{tot}/n_h$ of the $N$-pocket states in the hole density range where they are stable.
  At fixed hole density, the ground state is the one with the lowest energy.
  (b-d) Typical hole distributions of the $N$-pocket ground states.
  The shaded areas are given by the approximation Eq. \eqref{eq:non_dist_pocket} and the red lines are obtained from a full variational scheme, which reveal a significant degree of agreement.
  The hole density as taken as $n_h=0.06,0.16,0.27\times10^{12}\mathrm{cm}^{-2}$ respectively.
  }
  \label{fig:meta_stable_states}
\end{figure}

Under HF approximation, the total energy per area is expressed as a functional of the hole occupation function $n_{\bm{k}}\equiv 1-\ave{c^{\dagger}_{\bm{k}}c_{\bm{k}}}$ as $E_{tot}[n_{\bm{k}}]=E_{0}+E_{H}+E_{K}[n_{\bm{k}}]+E_{ex}[n_{\bm{k}}]$.
The vacuum energy of the fully filled HVB $E_0=\mathcal{S}^{-1}\sum_{\bm{k}}\varepsilon^0_{\bm{k}}$ is a constant.
Besides, we assume the system is charge uniform and the Hartree energy $E_H$ is also a constant.
Thus these two terms will be ignored in the following text.
The other two terms are kinetic energy $E_{K}$ and exchange energy $E_{ex}$, which are separately written as 
\begin{subequations}
  \begin{gather}
    E_{K}[n_{\bm{k}}]=\frac{1}{\mathcal{S}}\sum_{\bm{k}}(-\varepsilon^0_{\bm{k}})n_{\bm{k}}\label{eq:kin_energy}\\
    E_{ex}[n_{\bm{k}}]=-\frac{1}{2\mathcal{S}^2}\sum_{\bm{k}\bm{k}'}U_{\bm{k}\bm{k}'}(\bm{k}-\bm{k}')n_{\bm{k}}n_{\bm{k}'}\label{eq:ex_energy}
  \end{gather}\label{eq:tot_energy}
\end{subequations}
For very low doping concentration $n_h$ where the three hole Fermi surfaces as shown in Fig. \ref{fig:blg_non}(b) do not touch,
we can divide the Brillouin zone into three regions $\mathcal{Q}_{\alpha}$ ($\alpha=1,2,3$) as illustrated in Fig. \ref{fig:meta_stable_states}(b-d) and treat the mini-valley as a new degree of freedom.
Under this assumption, we further approximate the full occupation function $n_{\bm{k}}$ as three separate fermi-Dirac distributions for different mini-valleys which only depend on the corresponding valley occupation number $n_{\alpha}$. 
To be specific, the occupation function is approximated by
\begin{equation}
  n_{\bm{k}}\approx n^0_{\bm{k}}(n_{\alpha[\bm{k}]})=\Theta(\mu_{\alpha[\bm{k}]}+\varepsilon^0_{\bm{k}}), \label{eq:non_dist_pocket}
\end{equation}
where $\alpha[\bm{k}]$ is the mini-valley index of the wavevector $\bm{k}$ and $\Theta(\epsilon)$ is the Heaviside function (Fermi-Dirac distribution at zero temperature).
In Eq. \eqref{eq:non_dist_pocket}, the parameter $\mu_{\alpha}$ is the valley-dependent hole chemical potential determined from the hole concentration $n_{\alpha}$ by
\begin{equation}
  n_{\alpha}=\int_{\mathcal{Q}_{\alpha}}\frac{\rd^2k}{(2\pi)^2}\;n_{\bm{k}}=\int_{\mathcal{Q}_{\alpha}}\frac{\rd^2k}{(2\pi)^2}\;\Theta(\mu_{\alpha}+\varepsilon^0_{\bm{k}}).\label{eq:mu_n_implicit}
\end{equation}
Then we can treat the three $n_{\alpha}$s as the only variational parameters to greatly simplify the calculations. 
In Fig. \ref{fig:meta_stable_states}(b-d), the approximated hole occupation function gotten by Eq. \eqref{eq:non_dist_pocket} (shaded areas) are compared with those from the full variational scheme (red lines), which reveal a significant degree of agreement.
{Besides, the total energies calculated from the approximated occupation function Eq. \eqref{eq:non_dist_pocket} are also compared with the results from the full variational scheme in Appendix \ref{app:mean_field}, and the relative errors are in the order of $10^{-3}$.
These two facts indicate that Eq. \eqref{eq:non_dist_pocket} is a very good approximation of the occupation function.
}

Under this approximation, the total energy is expressed as a function of $n_{\alpha}$ as
\begin{equation}
  E_{tot}[n_{\bm{k}}]\approx E_K(n_{1},n_2,n_3)+E_{ex}(n_{1},n_2,n_3).\label{eq:total_energy_pcc}
\end{equation}
For numerically convenience, the exchange energy functional are separated into the intra- and inter-valley terms which are further approximated by Eq. \eqref{eq:exchange_intra_expansion} and Eq. \eqref{eq:exchange_inter_expansion} respectively.
By minimizing Eq. \eqref{eq:total_energy_pcc} with respect to $n_{\alpha}$ under constrain $n_1+n_2+n_3=n_h$, we find that the only possible stable sates (local minimums) are the $N$-pocket states ($N=1,2,3$), where there are $N$ hole pockets equally occupied.
And this is consistent with the previous study by \citet{dongIsospinMomentumpolarizedOrders2023}.
In Fig. \ref{fig:meta_stable_states}(a), the total energies per hole $E_{tot}/n_h$ of the $N$-pocket states are plotted in the hole density range where they are stable, and the ground state is the one with the lowest energy.  
The typical ground state hole distributions of the $N$-pocket states are also plotted in Fig. \ref{fig:meta_stable_states}(b-d).

\begin{figure}
  \centering
  \includegraphics[width=\linewidth]{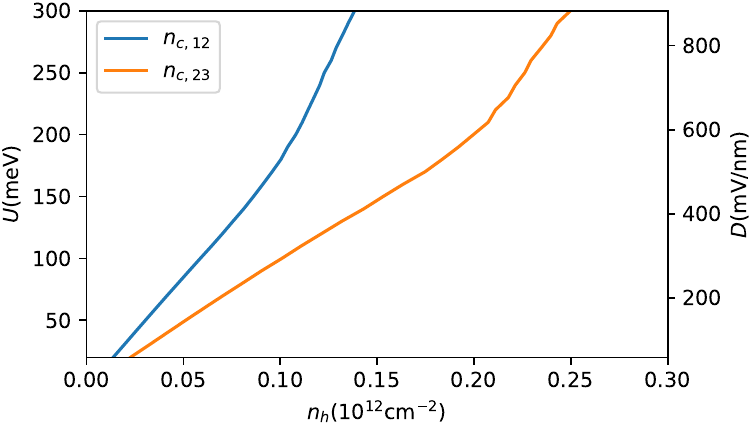}
  \caption{The critical density between the 1- and 2-pocket states $n_{c,12}$ and the critical density between the 2- and 3-pocket states $n_{c,23}$.
  }
  \label{fig:critical_density}
\end{figure}

{Denote the critical density between the 1- and 2-pocket states as $n_{c,12}$, the critical density between the 2- and 3-pocket states as $n_{c,23}$.
In Fig. \ref{fig:critical_density}, the critical densities are plotted as functions of the inter-layer bias potential $U$.
The corresponding displacement field $D=U/d$ ($d$ is the inter-layer distance) is also marked in the right axis for reference.
At $U=200\;\mathrm{meV}$ and $D\approx 600\;\mathrm{mV}/\mathrm{nm}$, the critical density for the 1- and 2-pocket state is about $n_{c,12}\approx 0.11\times 10^{12}\;\mathrm{cm}^{-2}$, which is a little larger than the experimentally result $0.07\times 10^{12}\;\mathrm{cm}^{-2}$ measured by \citet{linSpontaneousMomentumPolarization2023}.
The overestimation is reasonable since the screening effect of the interaction is not included in the Hartree Fock approximation.
}

\section{DC Transport: Valley Flipping and Current Hysteresis}\label{sec:dc}

\begin{figure*}
  \centering
  \includegraphics[width=0.9\linewidth]{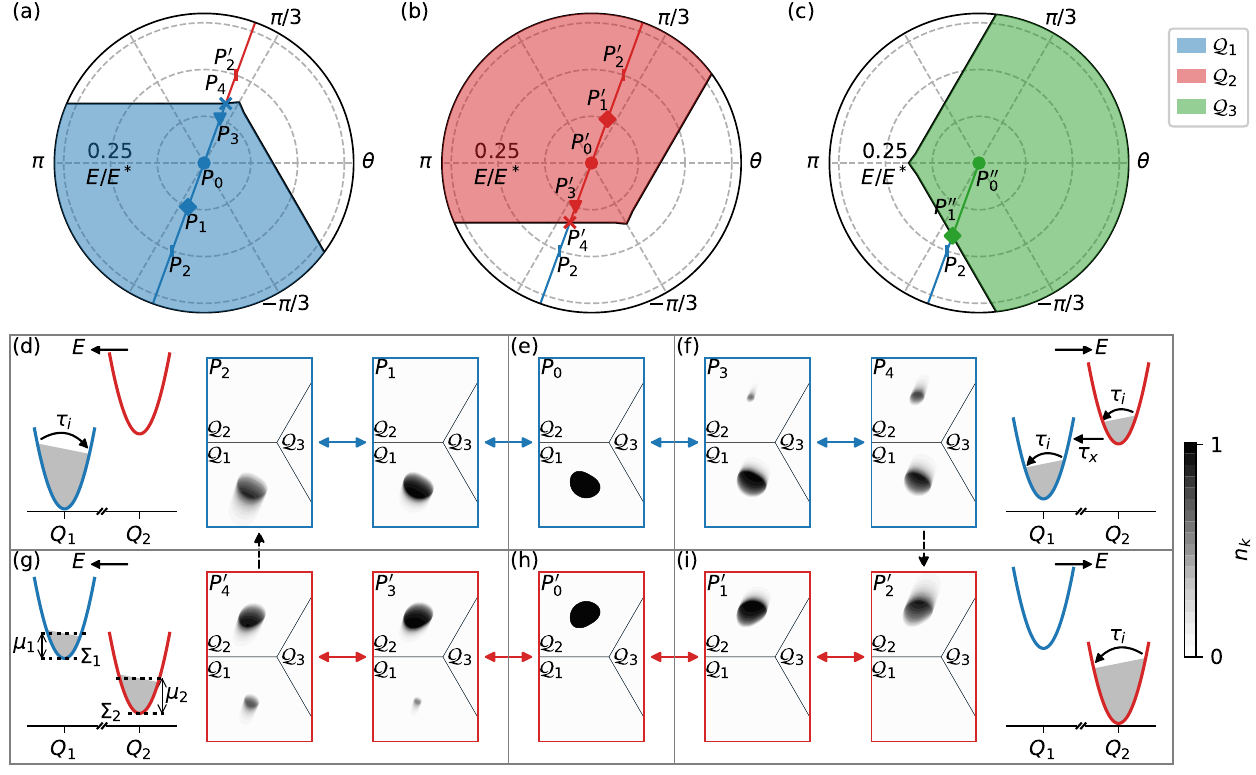}
  \caption{(a) Starting from the $\mathcal{Q}_1$-polarized state $P_0$ at zero electrical field strength, the steady hole distributions are smoothly connected if the electrical fields vary in the blue region and its boundary indicates the critical electrical field for valley flipping.
  (b)(c) The same quantities as (a) except that the initial states at zero field strength are polarized in the $\mathcal{Q}_2$ and $\mathcal{Q}_3$ valleys respectively.
  (d-f) Steady hole distributions for points $P_{0-4}$ in (a).
  (g-i) Steady hole distributions for points $P_{0-4}^{\prime}$ in (b).
  The blue and red color boxed panels in (d-i) form a hysteresis loop when the electrical field varies along the blue and red lines in (a) and (b).
  When generating these plots, the parameters are taken as $n_h=0.06\times10^{12}\;\mathrm{cm}^{-2}$, $\tau_i=100\;\mathrm{fs}$ and $\tau_x/\tau_i=100$.
  The field unit is defined as $E^*=\hbar\sqrt{3}Q/(2e\tau_i)\approx1.94\;\mathrm{V}/\mathrm{\mu m}$.
  The direction of the electrical field indicated by the blue and red lines is $\theta=7\pi/18$.
  }
  \label{fig:phase_dc}
\end{figure*}

In this section, we investigate the DC electrical transport properties of the valley polarized states.
For simplicity, we focus on the low hole density regime in this paper where only the $1$-pocket state is stable.
The Boltzmann equation for hole distribution function is
\begin{align}
  \frac{\partial n_{\bm{k}}}{\partial t}=&-\frac{e}{\hbar}\bm{E}\cdot\nabla_{\bm{k}}n_{\bm{k}}\nonumber\\
  &+\frac{2\pi n_{imp}}{\hbar}\int\frac{\rd^2k'}{(2\pi)^2}\;|T_{\bm{k}\bm{k}'}|^2\delta(\tilde{\varepsilon}_{\bm{k}}-\tilde{\varepsilon}_{\bm{k}'})(n_{\bm{k}'}-n_{\bm{k}}).\label{eq:boltzmann_initial}
\end{align}
The first term on the right side of the equation describes the driving effect by a DC electrical field and the second term is the impurity scattering, where $n_{imp}$ is the impurity concentration, $T_{\bm{k}\bm{k}'}$ is the impurity scattering matrix and $\tilde{\varepsilon}_{\bm{k}}\equiv \partial_{n_{\bm{k}}}E_{tot}[n_{\bm{k}}]$ is the quasi-particle energy of the holes.
Considering Eq. \eqref{eq:tot_energy}\eqref{eq:non_dist_pocket}, the quai-particle energy in the $\mathcal{Q}_{\alpha}$ valley could be approximated as $\tilde{\varepsilon}_{\bm{k}}\approx-\varepsilon^0_{\bm{k}}+\Sigma_{\bm{k}}$ where $\Sigma_{\bm{k}}\approx-{\mathcal{S}}^{-1}\sum_{\bm{k}'}U_{\bm{k}\bm{k}'}(\bm{k}-\bm{k}')n^0_{\bm{k}'}$ is the self-energy of holes due to interaction {(More precisely, the self-energy should be $\Sigma_{\bm{k}}=-{\mathcal{S}}^{-1}\sum_{\bm{k}'}U_{\bm{k}\bm{k}'}(\bm{k}-\bm{k}')n_{\bm{k}'}$, and the solutions of the Boltzmann equation with/without this approximation are compared in appendix. \ref{app:bol_sol}, which only have little difference)}.
Near the mini-valley center, the self-energy could be viewed as constants $\Sigma_{\bm{k}}\approx\Sigma_{\alpha[\bm{k}]}$ for different valleys, where $\Sigma_{\alpha}$ is obtained by $\Sigma_{\alpha}\approx \partial_{n_{\alpha}}E_{ex}(n_1,n_2,n_3)$.
As for the scattering matrix, we also assume that $|T_{\bm{k}\bm{k}'}|$ to be constants $T_i$ and $T_x$ for the intra- and inter-valley scattering process, which satisfies $T_i\gg T_x$.
For charge impurity with long range impurity potential in real space, the scattering matrix drops quickly with respect to the transferred momentum $|\bm{k}-\bm{k}'|$, and $T_i\gg T_x$ wil be a good approximation.
Under the above assumption, the intra-valley relaxation process is considered to be much faster than the inter-valley process. 
Therefore, we can approximately separate the fast and slow processes by ignoring the inter-valley relaxation process when dealing with the intra-valley dynamics.
Assume the hole concentration in $\mathcal{Q}_{\alpha}$ valley is $n_{\alpha}$ at time $t$, by adopting the relaxation time approximation, the Boltzmann equation in $\mathcal{Q}_{\alpha}$ valley becomes 
\begin{equation}
  \frac{\partial n_{\bm{k}}}{\partial t}\approx-\frac{e}{\hbar}\bm{E}\cdot\nabla_{\bm{k}}n_{\bm{k}} -\frac{n_{\bm{k}}-n_{\bm{k}}^{0}(n_{\alpha})}{\tau_i},\label{eq:distribution_dynamics}
\end{equation}
where $n^0_{\bm{k}}(n_{\alpha})$ is the Fermi-Dirac distributions given by Eq. \eqref{eq:non_dist_pocket}, $\tau_i$ is the intra-valley relaxation time\cite{mahanManyParticlePhysics2000} defined as $\tau_i^{-1}\equiv 2\pi n_{imp}|T_i|^2g/\hbar$ and $g$ is the DOS at the hole Fermi energy.
For 2D system, the DOS per mini-valley is nearly a constant and will be approximated by its value at the band edge which is about $g\approx 0.078 \;\mathrm{nm}^{-2}\mathrm{eV}^{-1}$ as shown in Fig. \ref{fig:blg_non}(e).
Using the Chamber's formula\cite{buddChamberSolutionBoltzmann1962}, Eq. \eqref{eq:distribution_dynamics} could be analytically solved as 
\begin{equation}
  n_{\bm{k}}(n_1,n_2,n_3)= \frac{1}{\tau_i}\int_{-\infty}^0\rd t'\re^{t'/\tau_i}n^0_{\bm{k}+e\bm{E}t'/\hbar}(n_{\alpha[\bm{k}+e\bm{E}t'/\hbar]})\label{eq:chamber}
\end{equation}

To further include the inter-valley process, we integrate Eq. \eqref{eq:boltzmann_initial} over $\mathcal{Q}_{\alpha}$ and the Boltzmann equation could be formally written as 
\begin{equation}
  \frac{\partial n_{\alpha}}{\partial t}=-\left(\frac{\partial n_{\alpha}}{\partial t}\right)_{d}+\left(\frac{\partial n_{\alpha}}{\partial t}\right)_{s}.\label{eq:pol_dynamics_formally}
\end{equation}
On the right side of Eq. \eqref{eq:pol_dynamics_formally}, the first term is the field driven term which is written as 
\begin{align}
  \left(\frac{\partial n_{\alpha}}{\partial t}\right)_{d}=&\int_{\mathcal{Q}_{\alpha}}\frac{\rd^2k}{(2\pi)^2}\;\frac{e}{\hbar}\bm{E}\cdot\nabla_{\bm{k}}n_{\bm{k}}(n_1,n_2,n_3)\nonumber\\
  =&\int_{\mathcal{Q}_{\alpha}}\frac{\rd^2k}{(2\pi)^2}\;\nabla_{\bm{k}}\cdot[\dot{\bm{k}}n_{\bm{k}}(n_1,n_2,n_3)].\label{eq:changing_rate_drive}
\end{align}
Notice that $\dot{\bm{k}}=e\bm{E}/\hbar$ is the velocity in $k$-space, we can define the current flows in $k$-space as $\bm{\mathcal{J}}_{\bm{k}}\equiv\dot{\bm{k}}n_{\bm{k}}$.
Using Gauss' law, Eq. \eqref{eq:changing_rate_drive} could be further written as 
\begin{equation}
  \left(\frac{\partial n_{\alpha}}{\partial t}\right)_{d}=\frac{1}{(2\pi)^2}\oint_{\partial \mathcal{Q}_{\alpha}}\bm{\mathcal{J}}_{\bm{k}}\cdot(\rd\bm{k}\times \hat{z}).
\end{equation}
In other words, the field driving term could be interpreted as changing rate of the particle numbers in area $\mathcal{Q}_{\alpha}$ caused by the $k$-space current flow $\bm{\mathcal{J}}_{\bm{k}}$.
The second term on the right side of Eq. \eqref{eq:pol_dynamics_formally} is the changing rate in area $\mathcal{Q}_{\alpha}$ due to impurity scattering.
According to the derivation in Appendix \ref{app:concentration_dynamics}, this term could be approximated by 
\begin{align}
  \left(\frac{\partial n_{\alpha}}{\partial t}\right)_{s}\approx &\frac{2\pi n_{imp}|T_{x}|^2}{\hbar} \sum_{\beta\ne \alpha}\left[\Theta(\Sigma_{\beta}+\mu_{\beta}-\Sigma_{\alpha}-\mu_{\alpha})\right.\nonumber\\
  &\left. \times g^2\min(\mu_{\beta},\Sigma_{\beta}+\mu_{\beta}-\Sigma_{\alpha}-\mu_{\alpha})-\alpha\leftrightarrow \beta\right],\label{eq:changing_rate_scatter}
\end{align}
As an example, we will use Fig. \ref{fig:phase_dc}(g) to explain the physical meaning of Eq. \eqref{eq:changing_rate_scatter}.
The inter-valley scattering from the $\mathcal{Q}_1$ valley to $\mathcal{Q}_2$ valley can only occur when $\mu_1+\Sigma_1>\mu_2+\Sigma_2$ is satisfied.
For a scattering channel with energy $\varepsilon$, the energy should satisfy $\Sigma_1<\varepsilon<\Sigma_1+\mu_1$ to make sure there are occupied initial states in $\mathcal{Q}_1$ valley.
Besides, the energy should also satisfy $\varepsilon>\Sigma_2+\mu_2$ to make sure there are empty final states in $\mathcal{Q}_2$ valley.
Thus the energy of the scattering channel should satisfy $\max(\Sigma_1,\Sigma_2+\mu_2)<\varepsilon<\Sigma_1+\mu_1$ and the total number of scattering channels are given by $g^2\min(\mu_1,\Sigma_1+\mu_1-\Sigma_2-\mu_2)$.
It's convenient to define the inter-valley relaxation time as $\tau_x^{-1}\equiv2\pi n_{imp}|T_x|^2g/\hbar$.
Then the ratio between these two relaxation times is $\tau_x/\tau_i=|T_i|^2/|T_x|^2\gg 1$, which means the intra-valley relaxation process is much faster than the inter-valley one.
By separating the fast relaxation process in the same mini-valley and the slow process across different valleys, the original Boltzmann equation Eq. \eqref{eq:boltzmann_initial} is decoupled into the two equations Eq. \eqref{eq:distribution_dynamics}\eqref{eq:pol_dynamics_formally}describing the intra and inter mini-valley dynamics respectively and will be the starting point of the further study.
From the above two equations, the steady state solution can be obtained under particular electric field, where the two terms on the right side of Eq. \eqref{eq:pol_dynamics_formally} cancel each other.

The 1-pocket state is threefold degenerate which could be polarized to $\mathcal{Q}_{\alpha=1,2,3}$ valleys with equal possibility.
As a consequence, the steady solution of the Boltzmann equation ($\partial_t n^S_{\bm{k}}=0$) is not only a function of the static electrical field, but also depends on the initial valley polarization.
For example, starting from the $\mathcal{Q}_1$-polarized state as illustrated by Fig. \ref{fig:phase_dc}(e) ($P_0$ point in Fig. \ref{fig:phase_dc}(a)), the final steady sates are evolved smoothly if the electrical fields only vary in the blue region in Fig. \ref{fig:phase_dc}(a) (the radial coordinate represents electrical field strength and the angular coordinate represents the direction).
When the electrical field is applied along the $\overrightarrow{P_0P_1P_2}$ direction in Fig. \ref{fig:phase_dc}(a), the holes will accelerate as $\hbar\dot{\bm{k}}=e\bm{E}$.
Such movements will be balanced by the intra-valley relaxation process, which results into a shifted steady hole distributions as shown in Fig. \ref{fig:phase_dc}(d) just like what happens in normal metals.
When the electrical field reverses to the $\overrightarrow{P_0P_3}$ direction before reaching the boundary of the blue region in Fig. \ref{fig:phase_dc}(a), holes will be driven into the $\mathcal{Q}_2$ valley which is balanced by the inter-valley relaxation process and the typical steady hole distributions are shown in Fig. \ref{fig:phase_dc}(f).
With the increase of the electrical field strength, the steady hole concentration in $\mathcal{Q}_2$ increases and there must be a critical point ($P_4$ in Fig. \ref{fig:phase_dc}(a)) where the inter-valley relaxation can not balance the field driving effect and a valley flipping occurs.
After that, the steady state will become the $\mathcal{Q}_2$-polarized state illustrated by Fig. \ref{fig:phase_dc}(h) ($P_0^{\prime}$ point in Fig. \ref{fig:phase_dc}(b)).
The cases for electrical fields varying along the red line in Fig. \ref{fig:phase_dc}(b) are similar and the corresponding steady hole distributions for points $P_{0-4}^{\prime}$ are shown in Fig. \ref{fig:phase_dc}(g-i).
Due to the valley flipping phenomenon, the steady hole distribution will form a hysteresis loop when electrical field strength takes values along the blue and red lines in Fig. \ref{fig:phase_dc}(a)(b).
Between red and blue boxed panels in Fig. \ref{fig:phase_dc}(d-i), the blue and red arrows indicate smooth reversible transitions while the black dashed arrow lines means irreversible.

\begin{figure}
  \centering
  \includegraphics[width=\linewidth]{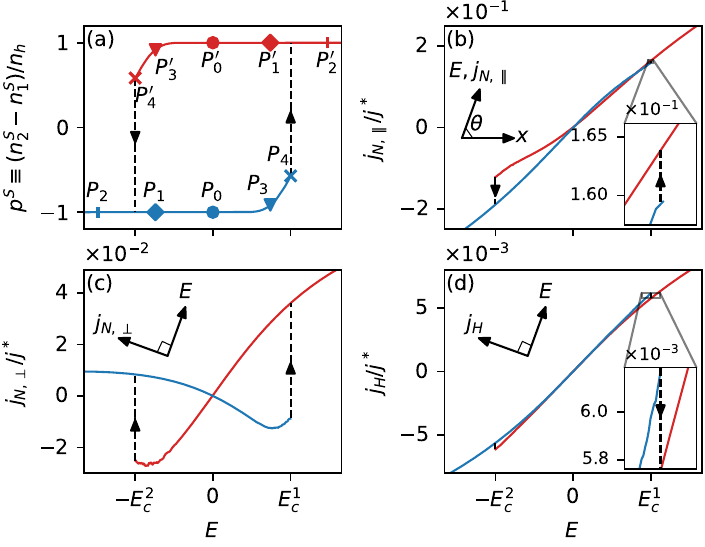}
  \caption{(a)(b)(c)(d) Hysteresis curves for $p^{S}$, $j_{N,\parallel}$ and $j_{N,\perp}$ and $j_{H}$ as functions of the electric field strength $E$ along the blue and red lines in Fig. \ref{fig:phase_dc}(a)(b).
  The critical field strengths for the valley flipping are about $E_{c}^{1}/E^*\approx 0.1693$ and $E_{c}^{2}/E^*\approx 0.1691$.
  The red and blue solid lines are reversible while the black dashed arrow lines are one-way through.
  The direction of the electrical field and the different current components are also indicated in the insets respectively.
  }
  \label{fig:hysteresis}
\end{figure}

To see the hysteretic effect more clearly, we define the polarization strength of the steady hole concentrations between $\mathcal{Q}_1$ and $\mathcal{Q}_2$ valleys as $p^{S}\equiv(n^{S}_2-n^{S}_1)/n_h$.
Then the hysteresis curve of $p^{S}$ along the blue and red lines in Fig. \ref{fig:phase_dc}(a)(b) is plotted in Fig. \ref{fig:hysteresis}(a).
The points $P_{0-4}$ and $P^{\prime}_{1-4}$ are also marked in Fig. \ref{fig:hysteresis}(a) accordingly.
The hysteresis of the distribution functions and hole concentrations will directly lead to the current hysteresis.
With the steady hole distributions $n_{\bm{k}}^{S}$ solved from the equations Eq. \eqref{eq:distribution_dynamics}\eqref{eq:pol_dynamics_formally}, the steady electrical current density is calculated as 
\begin{align}
  \bm{j}(n^{S}_{\bm{k}})=&\frac{e}{\hbar}\int\frac{\rd^2k}{(2\pi)^2}\;(\nabla_{\bm{k}}\tilde{\varepsilon}_{\bm{k}}+e \bm{E}\times\hat{z}\Omega_{\bm{k}})n^{S}_{\bm{k}}\nonumber\\
  \approx&\frac{e}{\hbar}\int\frac{\rd^2k}{(2\pi)^2}\;(-\nabla_{\bm{k}}{\varepsilon}^0_{\bm{k}}+e\bm{E}\times\hat{z}\Omega_{\bm{k}})n^{S}_{\bm{k}}.\label{eq:current_static}
\end{align}
In the above expression, the first term is the normal current $\bm{j}_{N}$ and the second term is the anomalous Hall current $\bm{j}_{H}$ due to the non-zero Berry curvature $\Omega_{\bm{k}}$ of HVB\cite{sundaramWavepacketDynamicsSlowly1999,xiao2010}.
Decomposed to the longitudinal direction of the electrical field $\hat{r}$ and transverse direction $\hat{\theta}\equiv\hat{z}\times \hat{r}$, the normal current has both longitudinal and transverse components $\bm{j}_{N}=j_{N,\parallel}\hat{r}+j_{N,\perp}\hat{\theta}$, while the Hall current is always along the transverse direction $\bm{j}_{H}=j_{H}\hat{\theta}$.
In Fig. \ref{fig:hysteresis}(b-d), the hysteresis curves for the different current components $j_{N,\parallel}$, $j_{N,\perp}$ and $j_{H}$ are also plotted.

In Fig. \ref{fig:phase_dc} and \ref{fig:hysteresis}, the hole density is taken as $n_h=0.06\times10^{12}\;\mathrm{cm}^{-2}$ and the relaxation times are taken as $\tau_i=100\;\mathrm{fs}$\citep{monteverdeTransportElasticScattering2010} and $\tau_x/\tau_i=100$.
The electrical field unit is defined as $E^*=\hbar\sqrt{3}Q/(2e\tau_i)\approx1.94\;\mathrm{V}/\mathrm{\mu m}$, where $Q=|\bm{Q}_{\alpha}-\bm{K}_{+}|$.
The current unit is defined by $j^*=e^2\tau_in_hE^*/\sqrt{m_{r}m_{\varphi}}\approx 3.33\;\mathrm{nA}/\mathrm{nm}$ where $m_{r}$ and $m_{\varphi}$ are effective masses at the mini-valley center in radial and angular direction respectively.

\section{AC Transport: Valley Oscillation And High-Order Nonlinearity}\label{sec:ac}

In this section, we replace the DC field discussed in last section by a sinusoidal AC field $\tilde{\bm{E}}(t)=\tilde{E}(\hat{x}\cos\theta+\hat{y}\sin\theta)\sin(2\pi t/T)$, which is more close to the realistic experimental setups.
We assume that the period $T$ of the external field is much longer than the inter-valley relaxation time $\tau_x$, which is always true for the experimental measurements below MHz.
Then an adiabatic approximation could be safely implemented.

\begin{figure}[h]
  \centering
  \includegraphics[width=\linewidth]{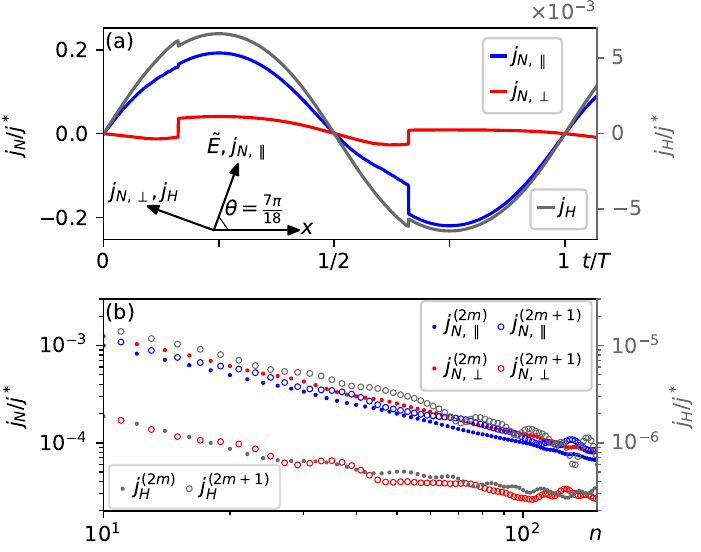}
  \caption{(a) Current responses in the time domain.
  The inset shows the direction of the AC electrical field and different current components.
  The longitudinal and transverse normal currents $j_{N,\parallel}$, $j_{N,\perp}$ (left axis) are encoded by blue and red colors respectively, while the Hall current $j_{H}$ (right axis) is represented by gray color.
  The field amplitude is taken as $\tilde{E}/E^*=0.2$, and the steady state is an oscillating state between $\mathcal{Q}_1$ and $\mathcal{Q}_2$ valleys.
  (b) The corresponding current responses in the frequency domain.
  The results for even ($n=2m$) and odd ($n=2m+1$) frequencies are represented by dots and circles respectively.
  Both the horizontal and vertical axes are drawn on logarithmic scales.
  }
  \label{fig:currents}
\end{figure} 

From the hysteresis curve in Fig. \ref{fig:hysteresis}(a), it's easy to see that the final state is an oscillating state between $\mathcal{Q}_1$ and $\mathcal{Q}_2$ valleys if the AC electrical field amplitudes satisfies $\tilde{E}>E_{c}^{1,2}$.
Taking $\tilde{E}/E^*=0.2$ and $\theta=7\pi/18$, the current responses during one period are directly got by tracing the hysteresis curves in Fig. \ref{fig:hysteresis}(b-d) and plotted in Fig. \ref{fig:currents}(a).
Due to the valley flipping phenomenon, the current responses in the time domain show discontinuous features when the electrical field passes the critical values.
By defining the current response at the $n$-th harmonic frequency as
\begin{equation}
  j^{(n)}\equiv\frac{1}{T}\int_0^{T}\rd t\;j(t)\re^{in\omega t},
\end{equation}
the current responses in the frequency domain are also calculated and plotted in Fig. \ref{fig:currents}(b).
As a result of the discontinuities in the time domain, the current responses in the frequency domain decay in a power-law manner, i.e. $\log j^{(n)}\sim -a \log n+b$ and $j^{(n)}\sim n^{-a}$.
This power-law decay leads to a very high-order nonlinearity, which is beyond the perturbation description.
For the perfect step-like discontinuities shown in Fig. \ref{fig:currents}(a), the power exponent is fitted as $a=1$.
While in realistic experiments, a larger exponent $a>1$ is expected, since the discontinuities are always smoothed out by inhomogeneity of the sample.

\begin{figure}[h]
  \centering
  \includegraphics[width=\linewidth]{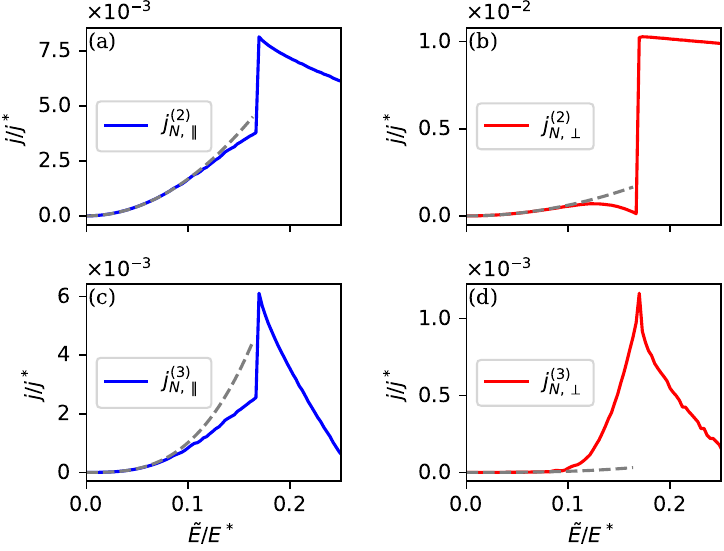}
  \caption{(a)(b) I-V curves of the second-harmonic responses. The dashed lines are quadratic fits.
  (c)(d) I-V curves of the third-harmonic responses. The dashed lines are cubic fits.
  }
  \label{fig:shg_thg}
\end{figure}

{
Such non-perturbative behaviors also appear in the I-V curves
of the high-harmonic nonlinear responses.
In Fig .\ref{fig:shg_thg}(a)(b), the second-harmonic current responses $j^{(2)}$ are plotted as functions of the AC electrical field amplitude $\tilde{E}$.
At small electrical field, the currents are well fitted by quadratic functions.
However, with the increase of the electrical field amplitude, the I-V curves gradually deviate from the quadratic fits, indicating the failure of the perturbation theory.
This agree qualitatively with the experimental results in Ref. \citep{linSpontaneousMomentumPolarization2023}(FIG. S1).
Besides, we also plot the I-V curves of the third-harmonic responses in Fig. \ref{fig:shg_thg}(c)(d), which show similar behaviors.
In spite of the deviation from the perturbation theory, we also find that these I-V curves are discontinuous at the critical electrical field strength.
This will help to determine the threshold field experimentally. 
}

\begin{figure}
  \centering
  \includegraphics[width=0.9\linewidth]{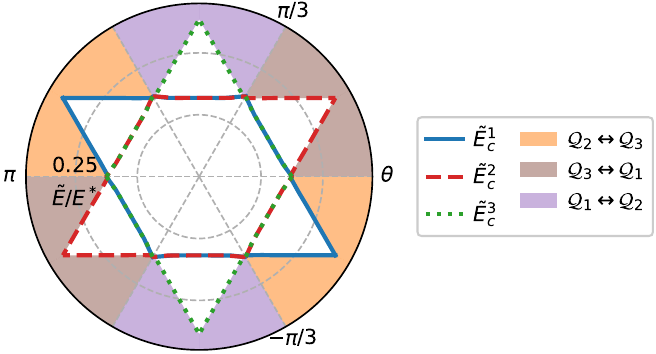}
  \caption{The line plots represent the critical AC electrical field amplitudes $\tilde{E}_c^{\alpha}$ for reaching a valley-oscillating state, where the superscripts $\alpha=1,2,3$ indicate the initial valley polarization $\mathcal{Q}_{\alpha}$.
  The filled regions represent the valley-oscillating states which are independent of the initial polarization.
  For example, the purple colored region labeled by $\mathcal{Q}_1\leftrightarrow\mathcal{Q}_2$ means that the final state is an oscillating state between $\mathcal{Q}_1$ and $\mathcal{Q}_2$ valleys when the electrical field amplitude is taken in this region.
  The maximum value of all the critical field amplitude is $\tilde{E}_M\equiv\max(\tilde{E}^{\alpha}_c)\approx 0.3177E^*$ and the minimum value is $\tilde{E}_m\equiv\min(\tilde{E}^{\alpha}_c)\approx 0.1589E^*$
  }
  \label{fig:phase_ac}
\end{figure}

If Fig. \ref{fig:phase_ac}, we also plot the critical amplitudes $\tilde{E}_c^{\alpha}$ for reaching the valley-oscillating states (the superscripts $\alpha=1,2,3$ indicate the initial valley polarization $\mathcal{Q}_{\alpha}$).
An interestingly condition is when the the electrical amplitude is larger than all of the three critical amplitudes $\tilde{E}>\tilde{E}_c^{\alpha=1,2,3}$.
If this is satisfied, the final valley-oscillating state will be independent of the initial polarization.
Using the oscillating state between $\mathcal{Q}_1$ and $\mathcal{Q}_2$ for example (the purple region labeled by $\mathcal{Q}_1\leftrightarrow\mathcal{Q}_2$ in Fig. \ref{fig:phase_ac}), it's not surprising to reach such an oscillating state if the initial state is polarized in $\mathcal{Q}_1$ or $\mathcal{Q}_2$ valleys as indicated by the hysteresis loop Fig. \ref{fig:phase_dc}(d-i) and Fig. \ref{fig:hysteresis}(a).
If the initial state is polarized in the $\mathcal{Q}_3$ valley, the system will be first flipped into the $\mathcal{Q}_1$-polarized state if the electrical field is applied along the line $\overrightarrow{P_0^{\prime\prime}P_1^{\prime\prime}P_2}$ in Fig. \ref{fig:phase_dc}(c).
After that, the system also enters into a steady oscillating state between the $\mathcal{Q}_1$ and $\mathcal{Q}_2$ valleys.

\begin{figure}
  \centering
  \includegraphics[width=\linewidth]{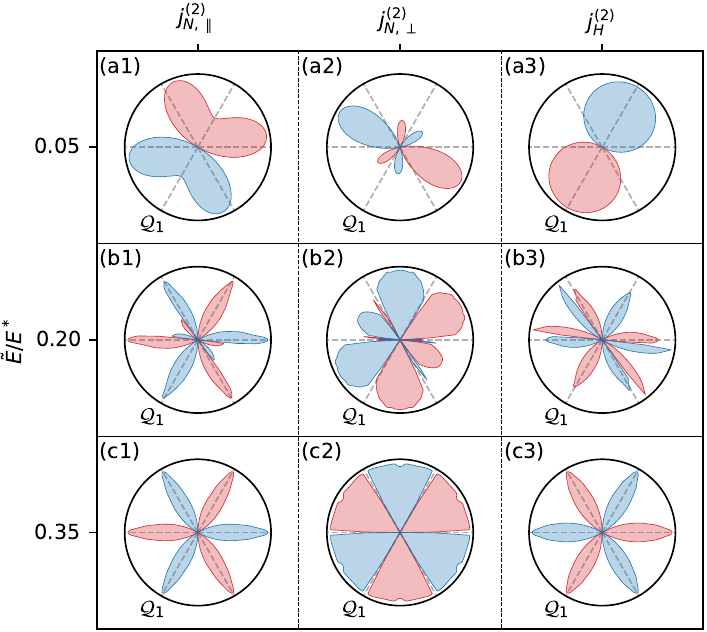}
  \caption{Current responses at the second harmonic frequency $j^{(2)}$ as functions of electrical field direction $\theta$ in polar coordinates.
  According to the sign of $\mathrm{Re}(j^{(2)})$, the plots are colored red and blue respectively.
  The initial valley polarization is also marked by the $\mathcal{Q}_1$ labels.
  Different rows mean different AC electrical field amplitudes while different columns represent different current components.
  }
  \label{fig:HHGs}
\end{figure}

The critical amplitudes in Fig. \ref{fig:phase_ac} have both maximum value $\tilde{E}_{M}=\max(\tilde{E}_c^{\alpha})\approx0.3177E^*$ and minimum value $\tilde{E}_{m}=\min(\tilde{E}_c^{\alpha})\approx0.1589E^*$ and the angular patterns of the current responses will behave differently for $\tilde{E}<\tilde{E}_m$ and $\tilde{E}>\tilde{E}_M$.
By assuming a $\mathcal{Q}_1$-polarized initial state, the second harmonic currents $j^{(2)}$ of the final periodic steady state are calculated and their norms $|j^{(2)}|$ are plotted in Fig. \ref{fig:HHGs} as functions of $\theta$ in polar coordinates.
Different rows mean different AC electrical field amplitudes as indicated in the left axis while different columns represent different current components as indicated in the top axis.
These plots are colored red and blue according to the sign of their real part $\mathrm{Re}(j^{(2)})$.
All the plots in Fig. \ref{fig:HHGs} are antisymmetric about the inversion operation, this arises from the additional tempo-spatial symmetry of the sinusoidal AC electrical field $\tilde{\bm{E}}(t;\theta+\pi)=\tilde{\bm{E}}(t+T/2;\theta)$ as discussed in Appendix \ref{app:symmetry}.
In the first row, the field amplitude is taken as $\tilde{E}=0.05E^*<\tilde{E}_{m}$ and no valley flipping occurs.
Then the current responses show $C_3$-breaking patterns.
In the third row, a much larger electrical field amplitudes $\tilde{E}=0.35E^*>\tilde{E}_M$ is taken and the broken $C_3$ symmetry recovers.
This is not surprising since the final states are valley-oscillating states in any electrical field direction as indicated by Fig. \ref{fig:phase_ac} and the system will lose its memory about the initial polarization.
At the intermediate field amplitudes, for example $\tilde{E}=0.2E^*$ in the second row, the angular patterns are similar to the first row in the directions with no valley flipping and similar to the third row in the directions where valley oscillation occurs.

\section{Discussion}\label{sec:discussion}

When the spin and atomic valley degeneracy is lifted in the quarter-metal phase, the inversion and time reversal symmetries are both broken, which ensures the existence of the anomalous Hall currents as illustrated by the gray line in Fig. \ref{fig:currents}(a).
At this time, the approximated point group is $D_{3d}$ which contains the threefold rotations $C_3$ and mirror symmetries $M_{\theta}$ with respect to the mirror lines $\theta=0,\;\pm\pi/3$.
If the system is further condensed to the 1-pocket state, for example the $\mathcal{Q}_1$-polarized state, the rotation symmetries are broken and only one mirror symmetry $M_{\pi/3}$ survives.
As a consequence of the low symmetry, all the three electrical current components $j_{N,\parallel}$, $j_{N,\perp}$ and $j_{H}$ usually exist at the same time and they have both even and odd frequency responses in the frequency domain as shown in Fig. \ref{fig:currents}(b).
To distinguish the two kind of currents in the transverse direction, the surviving mirror symmetry could be utilized.
According to Eq. \eqref{eq:mirror_app} in Appendix \ref{app:symmetry}, the normal current $j_{N,\perp}$ and Hall current $j_{H}$ have odd and even parities respectively under mirror operation.
And these symmetry restrictions are clearly shown in Fig. \ref{fig:HHGs} by the last two columns.

\begin{figure}[h]
  \centering
  \includegraphics[width=\linewidth]{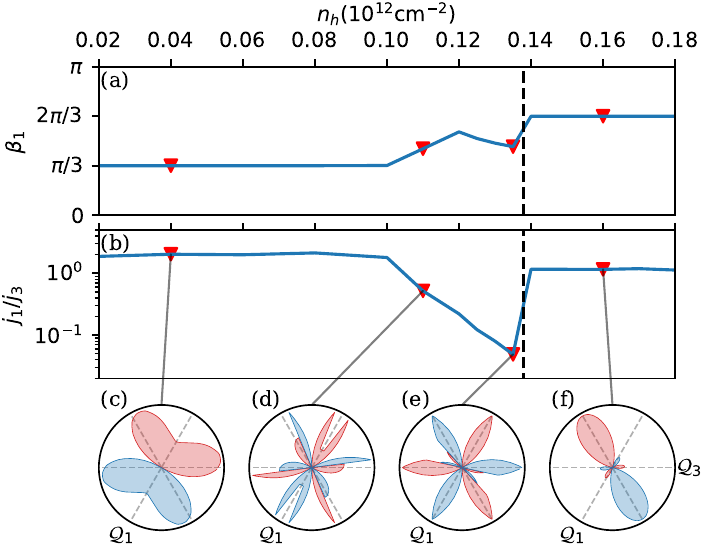}
  \caption{(a)(b) Polarization direction $\beta_1$ and polarization strength $j_1/j_3$ of the second order longitudinal current responses.
  The black dashed lines mark the critical density between the 1- and 2-pocket phases $n_{c,12}$.
  (c-f) Angular patterns of the second order longitudinal current responses.
  The corresponding hole densities are marked in (a)(b) by the red triangle symbols.
  The labels $\mathcal{Q}_{\alpha}$ marks the initial valley polarization direction, i.e., in (c-e) the initial state is a 1-pocket state polarized in $\mathcal{Q}_1$ valley, in (f) the initial state is a 2-pocket state polarized in $\mathcal{Q}_1$ and $\mathcal{Q}_3$ valleys.}
  \label{fig:pattern_density}
\end{figure}
In Ref. \citep{linSpontaneousMomentumPolarization2023}, the symmetry of the angular pattern of the second order longitudinal resistivity is used to determine the ground state.
However, as illustrated in Fig. \ref{fig:HHGs}(c1-c3), when the electrical field strength is strong enough to flip the valley polarization, the angular patterns may not reflect the ground state symmetry faithfully.
As discussed in Appendix \ref{app:Ec_nh_tau}, the critical field strength for valley flipping decreases with hole concentration $n_h$, suggesting a $C_3$-persevering pattern is likely to appear near the phase boundary between the 1- and 2-pocket phases.
Following Ref. \citep{linSpontaneousMomentumPolarization2023}, the second order longitudinal current responses as function of $\theta$ could be fitted by 
\begin{equation}
  j_{N,\parallel}^{(2)}(\theta)=j_1\cos(\theta-\beta_1)+j_3\cos(3(\theta-\beta_3)),
\end{equation} 
where $j_1/j_3$ measures the strength of polarization and $\beta_1$ indicates the polarization direction.
At fixed electrical field amplitude $\tilde{E}=0.125 E^*$, quantities $\beta_1$ and $j_1/j_3$ are plotted in Fig. \ref{fig:pattern_density}(a)(b) as functions of hole density $n_h$.
When there is no valley flipping in the low density regime, the angular pattern of $j^{(2)}_{N,\parallel}$ does reveal the symmetry of the initial $\mathcal{Q}_1$-polarized state as shown in Fig. \ref{fig:pattern_density}(c).
However, with the increase of the hole density $n_h$, due to the valley flipping, the polarization strength of the current responses $j_1/j_3$ decreases as shown in Fig. \ref{fig:pattern_density}(b), and the $C_3$ symmetry emerges gradually as illustrated by Fig. \ref{fig:pattern_density}(d)(e).
The dip of $j_1/j_3$ and a $C_3$-preserving pattern near the boundary of 1-pocket phase are both observed in Ref. \citep{linSpontaneousMomentumPolarization2023}(FIG.2), but might be wrongly interpreted as an unpolarized ground state in their paper.
When the hole density is larger than $n_{c,12}$, the ground state is a 2-pocket state and the emerged $C_3$ symmetry no longer exists as shown by Fig. \ref{fig:pattern_density}(f).
For better comparison, in Appendix \ref{app:sh_pattern}, we also study the angular pattern of the longitudinal resistivity which is directly measured in the experiment.
When the gating potential $U$ is small, the angular pattern of the second order longitudinal resistivity Fig. \ref{fig:pattern_potential}(b) agrees quantitatively with the experimental results in Ref. \citep{linSpontaneousMomentumPolarization2023}(FIG.2.b).

A concerning problem is spatial domains where the holes populate different mini-valleys, which make the system nonuniform.
However, as shown in our paper, the valley polarizations in different domains could be aligned by a large enough electrical field or equivalently current flow.
And the discussion of the uniform case in our paper is sufficient.
{Besides, it's also known that the free carriers are very easily localized in the presence of disorders in 2D. 
However, the experimental results in Ref. \citep{linSpontaneousMomentumPolarization2023} show that the devices are clean enough and there are no evidences signaling the localization behavior at the parameter regime we are interested in.
In such clean devices, one may worry that the system falls into the hydrodynamic transport regime, which requires the mean free path due to electron-electron scattering $l_e$ and impurity scattering $l_i$ satisfy $l_e\ll l_i$, or equivalently, the lifetime due to electron-electron scattering $\tau_e$ and impurity scattering $\tau_i$ satisfy $\tau_e\ll \tau_i$.
For 2D electron gases, the inverse lifetime due to electron-electron scattering scales along with the temperature $T$ as $\hbar/(\tau_e E_F)\sim -(k_B T/E_F)^2\ln(k_B T/E_F)$\citep{qianLifetimeQuasiparticleElectron2005}, where $E_F$ is the Fermi energy measured from the band minimum (for electrons) or maximum (for holes).
In \citet{linSpontaneousMomentumPolarization2023}, the measurements are conducted as $T=20\;\mathrm{mK}$; and at the hole density $n_h=0.1\times 10^{-12}\;\mathrm{cm}^{-2}$, the Fermi energy is about $10\;\mathrm{meV}$ as shown in Fig. \ref{fig:blg_non}(e).
Thus, the electron-electron scattering lifetime is estimated as $\tau_e\sim 10^{-7}\;\mathrm{s}$, which is much longer than the impurity lifetime $\tau_i$ in BLG (usually in the order of $10^{-14}\sim 10^{-12}\;\mathrm{s}$\citep{monteverdeTransportElasticScattering2010}).
This means that at the low temperature, the requirement for hydrodynamic transport couldn't be satisfied.
}

{The valley polarized phases and the related valley flipping phenomenon also have the potential for practical applications.}
In the AC case, when the driving field is strong enough, the power-law decay of the current responses in frequency domain as shown by Fig. \ref{fig:currents}(b) might be served as a high harmonic generators.
Additionally, as an ordered state which could be manipulated by electrical field and detected by transport measurements, the mini-valley polarized state holds promise for applications in information storage.

\begin{acknowledgments}

  We thank Prof. Guangyu Zhang for their inspiring experimental observations and helpful discussions.
  This work was fully supported by a fellowship award from the Research Grants Council of the Hong Kong Special Administrative Region, China (Project No. C7037-22GF)

\end{acknowledgments}

\appendix

{
\section{The Mean-field Approximation}\label{app:mean_field}
The total energy functional Eq. \eqref{eq:tot_energy} could be minimized be the following self-consistent equations 
\begin{subequations}
  \begin{gather}
    \tilde{\varepsilon}_{\bm{k}}=\frac{\delta E_{tot}}{\delta n_{\bm{k}}}=-\varepsilon_{\bm{k}}^0+\Sigma_{\bm{k}}[n_{\bm{k}}],\label{eq:quasiparticle_energy}\\
    \Sigma_{\bm{k}}[n_{\bm{k}}]=\frac{\delta E_{ex}}{\delta n_{\bm{k}}}=-\frac{1}{\mathcal{S}}\sum_{\bm{k}'}U_{\bm{k}\bm{k}'}(\bm{k}-\bm{k}')n_{\bm{k}'},\label{eq:self_energy}\\
    n_{h}=\frac{1}{\mathcal{S}}\sum_{\bm{k}}\Theta(\mu-\tilde{\varepsilon}_{\bm{k}}),\label{eq:chemical_potential}\\
    n_{\bm{k}}=\Theta(\mu-\tilde{\varepsilon}_{\bm{k}}).\label{eq:new_occupation_function}
  \end{gather}
\end{subequations}
To be specific, given an occupation function $n_{\bm{k}}^{l}$ (the superscript $l$ denotes the number of iteration loop), the quasi-particle energy and self-energy are given be Eq. \eqref{eq:quasiparticle_energy}, \eqref{eq:self_energy} respectively.
Then the overall chemical potential $\mu$ is determined by solving the equation Eq. \eqref{eq:chemical_potential} and the new distribution function $n_{\bm{k}}^{l+1}$ is given by Eq. \eqref{eq:new_occupation_function}.

\begin{figure}[h]
  \centering
  \includegraphics[width=\linewidth]{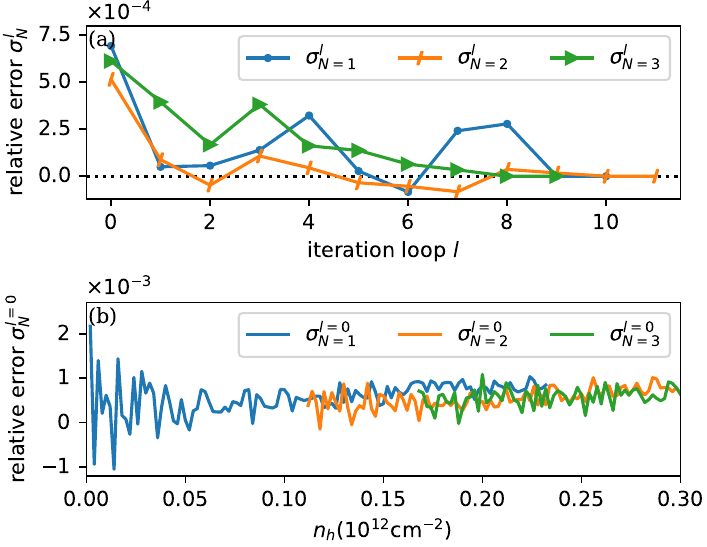}
  \caption{(a) Relative errors of the total energy as functions of iteration loop number $l$. For the $1$, $2$ and $3-$pocket states, the hole densities are taken as $n_h=0.06,0.16,0.27\times 10^{12}\;\mathrm{cm}^{-2}$ respectively.
  (b) Relative errors of the total energies at the beginning of the self-consistent iteration $\sigma_{N}^{l=0}$ as functions of the hole density $n_h$.}
  \label{fig:relative_error}
\end{figure}

Using the approximated occupation function Eq. \eqref{eq:non_dist_pocket} as an initial guess $n_{\bm{k}}^{l=0}$ for the $N$-pocket states, and the total energies during the iterations are given by $E_{N}^{l}=E_{tot}[n_{\bm{k}}^{l}]$.
Denote the converged occupation function as $n_{\bm{k}}^{f}$ and define the relative error of the total energy for the $N$-pocket state at the $l$-th iteration loop as 
\begin{equation}
  \sigma_N^{l}=\frac{E_{tot}[n_{\bm{k}}^l]-E_{tot}[n_{\bm{k}}^f]}{E_{tot}[n_{\bm{k}}^f]}.
\end{equation}
At some typical hole densities, the relative errors for the $N$-pocket states are plotted as a function of the iteration loop $l$ in Fig. \ref{fig:relative_error}(a).
Besides, the relative errors of the total energies at the beginning of the iteration are also plotted in Fig. \ref{fig:relative_error}(b) as functions of hole density $n_h$.
We can see that the iterations converge quickly and the relative errors of the total energies are in the order of $10^{-3}$ even with the initial guess of the hole distribution function $n_{\bm{k}}^0$.
Thus the approximation given by Eq. \eqref{eq:non_dist_pocket} is adequate and barely compromise the accuracy of the mean-field results.

}

\section{The Exchange Energy Functional}\label{app:exchange_energy}
To minimize $E_{tot}(n_1,n_2,n_3)$ in the main text, the HF exchange energy functional is further approximated by its series expansions with respect to $n_{\alpha=1,2,3}$.
For convenience, we define the intra-valley exchange energy $E_{ex}^{i}$ and inter-valley exchange $E_{ex}^{x}$ as 
\begin{align}
  E_{ex}^{i}=&-\frac{1}{2\mathcal{S}^2}\sum_{\alpha}\sum_{\bm{k}\bm{k}'\in \mathcal{Q}_{\alpha}}U_{\bm{k}\bm{k}'}(\bm{k}-\bm{k}')n_{\bm{k}}n_{\bm{k}'}\nonumber\\
  =&-\frac{1}{2}\sum_{\alpha}\int\frac{\rd^2\delta k\rd^2\delta k'}{(2\pi)^4}\;|\inp{\bm{Q}_{\alpha}+\delta\bm{k}}{\bm{Q}_{\alpha}+\delta\bm{k}'}|^2\nonumber\\
  &\times V_{col}(\delta\bm{k}-\delta\bm{k}')n^0_{\alpha,\bm{Q}_{\alpha}+\delta\bm{k}}n^0_{\alpha,\bm{Q}_{\alpha}+\delta\bm{k}'}\label{eq:exchange_intra}\\
  E_{ex}^{x}=&-\frac{1}{2\mathcal{S}^2}\sum_{\alpha\ne \beta}\sum_{\bm{k}\in\mathcal{Q}_{\alpha},\bm{k}'\in \mathcal{Q}_{\beta}}U_{\bm{k}\bm{k}'}(\bm{k}-\bm{k}')n_{\bm{k}}n_{\bm{k}'}\nonumber\\
  =&-\frac{1}{2}\sum_{\alpha\ne\beta}\int\frac{\rd^2\delta k\rd^2\delta k'}{(2\pi)^4}\;|\inp{\bm{Q}_{\alpha}+\delta\bm{k}}{\bm{Q}_{\beta}+\delta\bm{k}'}|^2\nonumber\\
  &\times V_{col}(\delta\bm{k}-\delta\bm{k}'+\bm{Q}_{\alpha}-\bm{Q}_{\beta})n^0_{\alpha,\bm{Q}_{\alpha}+\delta\bm{k}}n^0_{\beta,\bm{Q}_{\beta}+\delta\bm{k}'}\label{eq:exchange_inter}
\end{align}
where $\delta k\ll Q_{\alpha}$ is measured from the mini-valley center $\bm{Q}_{\alpha}$.
Due to the $1/q$ divergence of the Coulomb interaction $V_{col}(q)$ in Eq. \eqref{eq:exchange_intra}, the lowest order term when expanding with respect to $n_{\alpha}$ is $n_{\alpha}^{3/2}$, which is similar to the conventional 2D electron gas\citep{rajagopalCorrelationsTwodimensionalElectron1977}.
Then up to $n_h^3$, the intra-valley exchange energy is expanded as 
\begin{equation}
  E_{ex}^{i}\approx -\sum_{\alpha}(A_{1}n_{\alpha}^{3/2}+A_2n_{\alpha}^{5/2})+\mathcal{O}(n_h^{7/2}).\label{eq:exchange_intra_expansion}
\end{equation}
As for the intra-valley term Eq. \eqref{eq:exchange_inter}, since $\delta k\ll Q_{\alpha}$ is satisfied in the low doping regime, and $\delta\bm{k}-\delta\bm{k}'+\bm{Q}_{\alpha}-\bm{Q}_{\beta}$ can never be zero, there will be no singularities in the integration and the expansion of $E_{ex}^{x}$ with respect to $n_{\alpha}$ is the conventional Taylor expansion.
Then up to $n_h^3$, the inter-valley exchange energy should have the form 
\begin{equation}
  E_{ex}^{x}\approx -\sum_{\alpha\ne\beta}(B_1 n_{\alpha}n_{\beta}+B_2n_{\alpha}^2n_{\beta})+\mathcal{O}(n_h^{4}).\label{eq:exchange_inter_expansion}
\end{equation}
When the inter-layer gating potential is taken as $U=300\; \mathrm{meV}$ in the main text, the expansion coefficients are fitted from the numerical results as $A_1=0.429\;\mathrm{eVnm}$, $A_2=-16.0\;\mathrm{eVnm^3}$, $B_1=0.366\;\mathrm{eVnm^2}$ and $B_2=41.1\;\mathrm{eVnm^4}$.

\section{The Inter-valley Scattering Term}\label{app:concentration_dynamics}
By integrating over the $\mathcal{Q}_{\alpha}$ valley, the changing rate of hole concentrations in area $\mathcal{Q}_{\alpha}$ due to impurity scattering $(\frac{\partial n_{\alpha}}{\partial t})_s$ is obtained as
\begin{align}
  &\frac{2\pi n_{imp}}{\hbar}\int_{\mathcal{Q}_{\alpha}}\frac{\rd^2 k}{(2\pi)^2}\int\frac{\rd^2k'}{(2\pi)^2}\;|T_{\bm{k}\bm{k}'}|^2\delta(\tilde{\varepsilon}_{\bm{k}}-\tilde{\varepsilon}_{\bm{k}'})(n_{\bm{k}'}-n_{\bm{k}})\nonumber\\
  =&\frac{2\pi n_{imp}|T_i|^2}{\hbar}\int_{\mathcal{Q}_{\alpha}}\frac{\rd^2k}{(2\pi)^2}\int_{\mathcal{Q}_{\alpha}}\frac{\rd^2k'}{(2\pi)^2}\;\delta(\tilde{\varepsilon}_{\bm{k}}-\tilde{\varepsilon}_{\bm{k}'})(n_{\bm{k}'}-n_{\bm{k}})\nonumber\\
  +&\frac{2\pi n_{imp}|T_x|^2}{\hbar}\int_{\mathcal{Q}_{\alpha}}\frac{\rd^2k}{(2\pi)^2}\int_{\bar{\mathcal{Q}}_{\alpha}}\frac{\rd^2k'}{(2\pi)^2}\;\delta(\tilde{\varepsilon}_{\bm{k}}-\tilde{\varepsilon}_{\bm{k}'})(n_{\bm{k}'}-n_{\bm{k}})
\end{align}
where $\bar{\mathcal{Q}}_{\alpha}$ means regions exclude $\mathcal{Q}_{\alpha}$.
The two terms describe the intra- and inter-valley scattering processes respectively.
The intra-valley term change sign when switching the indices $\bm{k}$ and $\bm{k}'$ which means the integral is zero.
This reflects the fact that the intra-valley scattering process will not effect the hole concentrations in each valley.
To further simplify the inter-valley term, we approximate the distribution $n_{\bm{k}}$ in the integrand by $n^0_{\bm{k}}$ and the last term becomes
\begin{align}
  &\frac{2\pi n_{imp}|T_x|^2}{\hbar}\int_{\mathcal{Q}_{\alpha}}\frac{\rd^2k}{(2\pi)^2}\int_{\bar{\mathcal{Q}}_{\alpha}}\frac{\rd^2k'}{(2\pi)^2}\;\delta(\tilde{\varepsilon}_{\bm{k}}-\tilde{\varepsilon}_{\bm{k}'})(n^0_{\bm{k}'}-n^0_{\bm{k}})\nonumber\\
\approx&\frac{2\pi n_{imp}|T_x|^2g^2}{\hbar}\sum_{\beta\ne \alpha}\int\rd\varepsilon\int\rd\varepsilon'\;\delta(\varepsilon+\Sigma_{\alpha}-\varepsilon'-\Sigma_{\beta})\nonumber\\
&\qquad\times[\Theta(\mu_{\beta}-\varepsilon')-\Theta(\mu_{\alpha}-\varepsilon)]\nonumber\\
  =&\frac{2\pi n_{imp}|T_{x}|^2g^2}{\hbar} \sum_{\beta\ne \alpha}\left[\Theta(\Sigma_{\beta}+\mu_{\beta}-\Sigma_{\alpha}-\mu_{\alpha})\right.\nonumber\\
  &\qquad\left. \times \min(\mu_{\beta},\Sigma_{\beta}+\mu_{\beta}-\Sigma_{\alpha}-\mu_{\alpha})-\alpha\leftrightarrow \beta\right].\label{eq:inter-scatter}
\end{align}

{
\section{Effect of Coulomb Interaction to the Solution of Boltzmann Equation}\label{app:bol_sol}
When the hole concentrations in each mini-valley are fixed as $(n_1,n_2,n_3)$, the solution of the Boltzmann equation Eq. \eqref{eq:distribution_dynamics} is approximately given by Eq. \eqref{eq:chamber} in the main text.
However, once the hole distribution function is deformed by the electrical field, the self-energy 
\begin{equation}
  \Sigma_{\bm{k}}[n_{\bm{k}}]=-\frac{1}{\mathcal{S}}\sum_{\bm{k}'}U_{\bm{k}\bm{k}'}n_{\bm{k}'}
\end{equation}
and the quasiparticle energy $\tilde{\varepsilon}_{\bm{k}}=-\varepsilon_{\bm{k}}^{0}+\Sigma_{\bm{k}}$ also changes accordingly.
Thus, to fully include the effect of the interaction, one should solve the Boltzmann equation Eq. \eqref{eq:distribution_dynamics} self-consistently.
And the self-consistent equation reads
\begin{subequations}
  \begin{gather}
    \tilde{\varepsilon}_{\bm{k}}=-\varepsilon_{\bm{k}}^0-\frac{1}{\mathcal{S}}\sum_{\bm{k}'}U_{\bm{k}\bm{k}'}n_{\bm{k}'}\label{eq:quasiparticle_energy_2}\\
    n_{\alpha}=\frac{1}{\mathcal{S}}\sum_{\bm{k}\in \mathcal{Q}_{\alpha}}\Theta(\mu_{\alpha[\bm{k}]}-\tilde{\varepsilon}_{\bm{k}})\label{eq:chemical_valley}\\
    n_{\bm{k}}=\frac{1}{\tau_i}\int_{-\infty}^{0}\rd t'\;\re^{t'/\tau_i}\Theta(\mu_{\alpha[\bm{k}+e\bm{E}t/\hbar]}-\tilde{\varepsilon}_{\bm{k}+e\bm{E}t/\hbar})\label{eq:dist_E}
  \end{gather}\label{eq:bol_scf}
\end{subequations}
Given distribution function $n_{\bm{k}}$, the quasiparticle energy is determined by Eq. \eqref{eq:quasiparticle_energy_2}.
Then the chemical potentials in each mini-valley $\mu_{\alpha[\bm{k}]}$ are determined by solving Eq. \eqref{eq:chemical_valley}.
Finally, a new distribution function is got from Eq. \eqref{eq:dist_E}.
Using the same parameters as in Fig. \ref{fig:hysteresis}, the hysteresis curve of the polarization strength $p^S$ is recalculated and plotted in Fig. \ref{fig:hysteresis_comparison} by the solid blue line.
For comparison, the results in the main text where the Boltzmann equation is not self-consistently solved are also plotted in Fig. \ref{fig:hysteresis_comparison} by the dotted red line.

\begin{figure}[h]
  \centering
  \includegraphics[width=\linewidth]{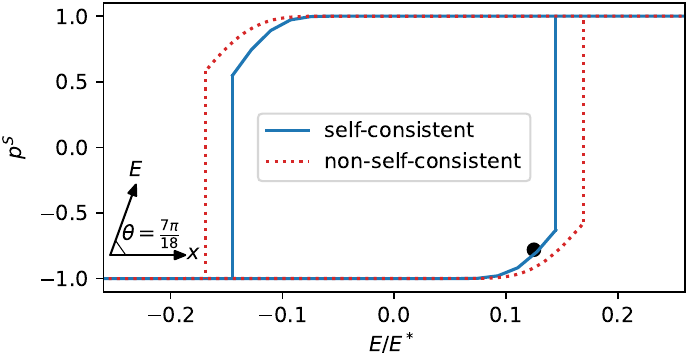}
  \caption{{The hysteresis curve of the polarization strength $p^{S}$. 
  When getting the solid blue line, the Boltzmann equation Eq. \eqref{eq:distribution_dynamics} is solved self-consistently according to Eq. \eqref{eq:bol_scf}.
  And the dotted red line represents the results where the self-energy is not determined self-consistently (the results in the main text).}
  }
  \label{fig:hysteresis_comparison}
\end{figure}

At $E/E^*=0.125$ (the black dot in Fig. \ref{fig:hysteresis_comparison}), the steady hole distribution function is calculated self-consistently and plotted in Fig. \ref{fig:dist_E_scf}(a).
Along the red dotted line in Fig. \ref{fig:dist_E_scf}(a), the quasiparticle energy and the hole distribution function are also plotted in Fig. \ref{fig:dist_E_scf}(b)(c).
In Fig. \ref{fig:dist_E_scf}(a), the original band minimums are marked by the three red cross symbols.
In the main text, the self-energy is approximated by $\Sigma_{\bm{k}}[n_{\bm{k}}^0]\approx\Sigma_{\bm{k}}(n_{\alpha[\bm{k}]})$, which will not change the location of band minimums
However, when the self-energy is determined self-consistently, we find that the minimums will also shift along the direction of the electrical field to the points marked by the blue cross symbols.

\begin{figure}[h]
  \centering
  \includegraphics[width=\linewidth]{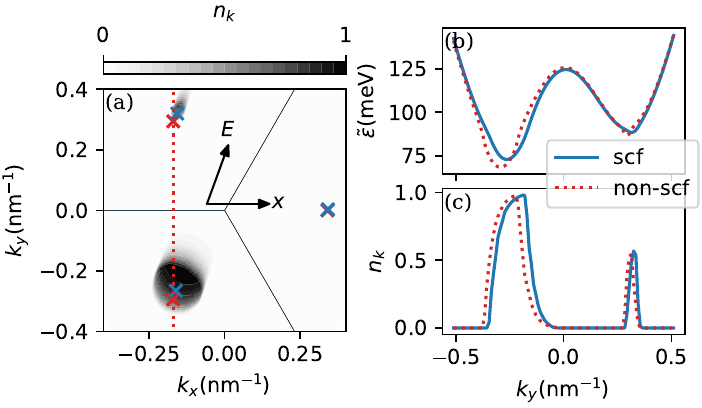}
  \caption{{(a) The steady hole distribution function. When generating this results, the Boltzmann equation Eq. \eqref{eq:distribution_dynamics} is solved self-consistently.
  The three red cross symbols mark the minimum of $-\varepsilon^0_{\bm{k}}$ and the three blue cross symbols mark the minimum of $\tilde{\varepsilon}_{\bm{k}}$ which is self-consistently determined from Eq. \eqref{eq:bol_scf}.
  (b)(c) The quasiparticle energy and the hole distribution function along the dotted red line in (a). 
  The solid blue lines and the dotted red lines represent the self-consistent and non-self-consistent results respectively.
  }}\label{fig:dist_E_scf}
\end{figure}

According to the comparisons above, when the self-energy is determined self-consistently rather than approximated by $\Sigma_{\bm{k}}[n^0_{\bm{k}}]$, the results indeed changes slightly.
But the physical pictures are not effected.

}
\section{Dependence of the Critical Field on Hole Concentration and Relaxation Times}\label{app:Ec_nh_tau}
In this part, we investigate the dependence of the critical field strength for valley flipping on hole concentration $n_h$ and relaxation times $\tau_{i,x}$.
Without losing generality, we will assume that the electrical field is applied in the direction $\pi/3<\theta<\pi$ and the initial state is $\mathcal{Q}_1$-polarized.
Other cases could by easily got by symmetry operations.

we first consider the zero-doping limit $n_h\to 0$.
At this time, the dynamics equation could be much simplified and analytically solved.
Define the driving coefficient as
\begin{align}
  \eta\equiv& \lim_{n_h\to 0}\frac{1}{n_1}\left(\frac{\partial n_1}{\partial t}\right)_d\nonumber\\
  =&\lim_{n_{h}\to 0}\frac{1}{n_1}\int_{\mathcal{Q}_1}\frac{\rd^2k}{(2\pi)^2}\;\frac{e}{\hbar}\bm{E}\cdot\nabla_{\bm{k}}n_{\bm{k}}\nonumber\\
  =&\lim_{n_{h}\to 0}\frac{1}{n_1}\int_{\mathcal{Q}_1}\frac{\rd^2k}{(2\pi)^2}\;\int_{-\infty}^0\rd t'\;\frac{\re^{t'/\tau_i}}{\tau_i}\frac{e}{\hbar}\bm{E}\cdot\nabla_{\bm{k}}n^0_{\bm{k}+e\bm{E}t'/\hbar}.\label{eq:driving_rate_def}
\end{align}
In the last line, the chamber formula Eq. \eqref{eq:chamber} is used.
According to the defining Eq. \eqref{eq:non_dist_pocket}, we have
\begin{align}
  \lim_{n_h\to 0}n^0_{\bm{k}}=&(2\pi)^2\sum_{\alpha}n_{\alpha}\delta(\bm{k}-\bm{Q}_{\alpha})\nonumber\\
  =&(2\pi)^2[n_1\delta(\bm{k}-\bm{Q}_1)+n_2\delta(\bm{k}-\bm{Q}_2)].\label{eq:zero_dist}
\end{align}
In Eq. \eqref{eq:zero_dist}, we only keep the contributions from the $\mathcal{Q}_{1}$ and $\mathcal{Q}_2$ valley.
This is because an electrical field in the direction $\pi/3<\theta<\pi$ can only drive holes from the $\mathcal{Q}_1$ to the $\mathcal{Q}_2$ valley and the $\mathcal{Q}_3$ valley is irrelevant, i.e. $n_3=0$. 
Notice that $e\bm{E}/\hbar\cdot \nabla_{\bm{k}}n^0_{\bm{k}+e\bm{E}t'/\hbar}=\partial_{t'}n^0_{\bm{k}+e\bm{E}t'/\hbar}$, Eq. \eqref{eq:driving_rate_def} becomes
\begin{align}
  \eta=&\frac{1}{n_1}\int_{-\infty}^{0}\rd t'\;\frac{\re^{t'/\tau_i}}{\tau_i}\partial_{t'}\left[\int_{\mathcal{Q}_1}\frac{\rd^2k}{(2\pi)^2}\;\lim_{n_h\to 0}n^0_{\bm{k}+e\bm{E}t'/\hbar}\right]\nonumber\\
  =&\frac{1}{n_1}\int_{-\infty}^{0}\rd t'\;\frac{\re^{t'/\tau_i}}{\tau_i}\partial_{t'}\int_{\mathcal{Q}_1}{\rd^2k}\;[n_1\delta(\bm{k}+e\bm{E}'t/\hbar-\bm{Q}_1)\nonumber\\
  &\qquad\qquad\qquad+n_2\delta(\bm{k}+e\bm{E}'t/\hbar-\bm{Q}_2)].\label{eq:driving_rate_2}
\end{align}
Since the electrical field is applied in the direction $\pi/3<\theta<\pi$, we always have $\bm{Q}_2-e\bm{E}t'/\hbar\in \mathcal{Q}_2$ for $t'<0$.
Thus 
\begin{equation}
  \int_{\mathcal{Q}_1}\rd^2k\;\delta(\bm{k}+e\bm{E}t'/\hbar-\bm{Q}_2)=0.\label{eq:Q2_integral}
\end{equation}
Besides, only when $t'>-\sqrt{3}\hbar Q/(2eE\sin\theta)$ is satisfied, the vector $\bm{Q}_1-e\bm{E}t'/\hbar$ is in $\mathcal{Q}_1$ valley.
Therefore
\begin{equation}
  \int_{\mathcal{Q}_1}\rd^2k\;\delta(\bm{k}+e\bm{E}t'/\hbar-\bm{Q}_1)=\Theta[t'+\sqrt{3}\hbar Q/(2eE\sin\theta)].\label{eq:Q1_integral}
\end{equation}
Using Eq. \eqref{eq:Q1_integral}\eqref{eq:Q2_integral}, the driving coefficient Eq. \eqref{eq:driving_rate_2} becomes
\begin{align}
  \eta=&\int_{-\infty}^{0}\rd t'\;\frac{\re^{t'/\tau_i}}{\tau_i}\partial_{t'}\Theta[t'+\sqrt{3}\hbar Q/(2eE\sin\theta)]\nonumber\\
  =&\frac{\re^{-\sqrt{3}\hbar Q/(2eE\tau_i\sin\theta)}}{\tau_i}.
\end{align}
In the zero-doping limit, the scattering term could also be simplified.
According to Eq. \eqref{eq:exchange_intra_expansion} and \eqref{eq:exchange_inter_expansion}, we have
\begin{equation}
  \lim_{n_h\to 0}\Sigma_{\alpha}=\lim_{n_h\to 0}\partial_{n_{\alpha}}E_{ex}=-3A_1 n_{\alpha}^{1/2}/2+\mathcal{O}(n_h^{3/2}).
\end{equation}
Besides, according to Eq. \eqref{eq:mu_n_implicit}, we also have 
\begin{equation}
  \lim_{n_h\to 0}\mu_{\alpha}=n_{\alpha}/g+\mathcal{O}(n_h^2).
\end{equation}
Thus the inter-valley scattering term becomes 
\begin{align}
  &\lim_{n_h\to 0}\frac{g}{\tau_x} \sum_{\beta\ne \alpha}\left[\Theta(\Sigma_{\beta}+\mu_{\beta}-\Sigma_{\alpha}-\mu_{\alpha})\right.\nonumber\\
  &\qquad\left. \times \min(\mu_{\beta},\Sigma_{\beta}+\mu_{\beta}-\Sigma_{\alpha}-\mu_{\alpha})-\alpha\leftrightarrow \beta\right]\nonumber\\
  =&\lim_{n_h\to 0}\frac{g}{\tau_x} \sum_{\beta\ne \alpha}\left[\Theta(\Sigma_{\beta}-\Sigma_{\alpha})\min(\mu_{\beta},\Sigma_{\beta}-\Sigma_{\alpha})-\alpha\leftrightarrow \beta\right]\nonumber\\
  =&\lim_{n_h\to 0}\frac{g}{\tau_x}\sum_{\beta\ne \alpha}[\Theta({n_{\alpha}}-{n_{\beta}})\mu_{\beta}-\alpha\leftrightarrow \beta]\nonumber\\
  =&\frac{1}{\tau_x}\sum_{\beta\ne \alpha}\left[\Theta(n_{\alpha}-n_{\beta})n_{\beta}-\alpha\leftrightarrow \beta\right].
\end{align}
In summary, the dynamics equation for $n_{\alpha}$s in the zero-doping limit is finally simplified to
\begin{gather*}
  \dot{n}_1=-\eta n_1+\tau_x^{-1}[\Theta(n_1-n_2)n_2-1\leftrightarrow 2],\\
  \dot{n}_2=\eta n_1+\tau_x^{-1}[\Theta(n_2-n_1)n_1-1\leftrightarrow 2].
\end{gather*}
Define the valley polarization strength as $p\equiv (n_2-n_1)/n_h$, then the dynamics equation of $p$ is 
\begin{align}
  \dot{p}=&\eta (1-p)+\tau_x^{-1}\left[\Theta(p)(1-p)-\Theta(-p)(1+p)\right]\nonumber\\
  =&(\eta+\tau_x^{-1}\mathrm{sgn}\;p)-(\eta+\tau_x^{-1})p
\end{align}
The steady polarization is gotten by $\dot{p}^{S}=0$, which gives 
\begin{subequations}
  \begin{gather}
    p^{S}=\frac{\eta-\tau_x^{-1}}{\eta+\tau_x^{-1}},\;p^{S}\le 0\;\\
    p^{S}=1,\;p^{S}>0.
  \end{gather}
\end{subequations}
The solution is discontinuous at $p^{S}=0$, which gives the critical electrical field for valley flipping as, 
\begin{equation}
  \frac{E_c}{E^*}=\frac{1}{\sin\theta\ln(\tau_x/\tau_i)}\label{eq:Ec_zero_doping}
\end{equation}
where $E^*=\sqrt{3}\hbar Q/(2e\tau_i)$ is the electrical field unit.

\begin{figure}
  \centering
  \includegraphics[width=\linewidth]{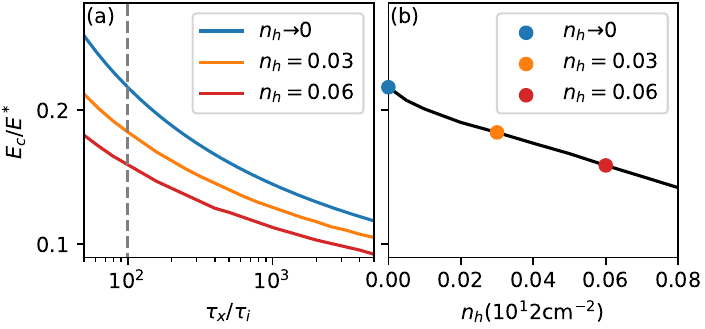}
  \caption{(a) Critical field for valley flipping as a function of $\tau_x/\tau_i$ at fixed electrical field direction $\theta=\pi/2$ and the initial state is $\mathcal{Q}_1$-polarized.
  The blue line for the zero doping limit is directly got by Eq. \eqref{eq:Ec_zero_doping} and the orange and red lines at finite doping concentrations are numerically solved from the Boltzmann equation.
  (b) The critical field strength as a function of hole concentrations at $\tau_x/\tau_i=100$ (the dashed gray line in (a)).}
  \label{fig:Ec_nh_tau}
\end{figure}

At $\theta=\pi/2$, the critical field strength in the zero doping limit Eq. \eqref{eq:Ec_zero_doping} is plotted as a function of $\tau_x/\tau_i$ in Fig. \ref{fig:Ec_nh_tau}(a) by the blue line.
Besides, numerically results by solving the Boltzmann equation at finite doping concentrations are also plotted for comparisons.
By taking the relaxation times as $\tau_x/\tau_i=100$ (the gray dashed in Fig. \ref{fig:Ec_nh_tau}(a)), the critical field strength is also plotted as a function of hole doping strength $n_h$ in Fig. \ref{fig:Ec_nh_tau}(b).

\section{Symmetry Restrictions on the Angular Pattern of Current Responses}\label{app:symmetry}
Assume the system is $\mathcal{Q}_1$-polarized and the surviving mirror symmetry is $M=M_{\pi/3}$ with respect to the mirror line $\theta=\pi/3$.
After the mirror operation, the electrical field and steady hole distribution transform to $M\bm{E}$ and $n^{S}_{M\bm{k}}$ respectively, and the current becomes
\begin{align}
  \bm{j}'=&\frac{e}{\hbar}\int\frac{\rd^2k}{(2\pi)^2}\;[-\nabla_{\bm{k}}{\varepsilon}^0_{\bm{k}}+e(M\bm{E})\times\hat{z}\Omega_{\bm{k}}]n^{S}_{M\bm{k}}\nonumber\\
  =&\frac{e}{\hbar}\int\frac{\rd^2k}{(2\pi)^2}\;M[-\nabla_{\bm{k}}{\varepsilon}^0_{\bm{k}}+e(M\bm{E})\times\hat{z}\Omega_{\bm{k}}]n^{S}_{\bm{k}}\nonumber\\
  =&M\bm{j}_{N}-\bm{j}_H.
\end{align}
We can see that the normal current $\bm{j}'_{N}=M\bm{j}_{N}$ transforms along with the mirror operation while the Hall current $\bm{j}'_{H}=-\bm{j}_{H}$ only changes sign.
This reflects the fact that the Hall current is a pseudo-vector.
Besides, after the mirror operation, the unit vectors in longitudinal and transverse directions also change to $\hat{r}'=M\hat{r}$ and $\hat{\theta}'=\hat{z}\times(M\hat{r})=-M(\hat{z}\times\hat{r})=-M\hat{\theta}$. 
Thus the different current components transform as $j'_{N,\parallel}\equiv \hat{r}'\cdot \bm{j}'_{N}=j_{N,\parallel}$, $j'_{N,\perp}\equiv \hat{\theta}'\cdot \bm{j}'_{N}=-j_{N,\perp}$ and $j'_{H}\equiv\hat{\theta}'\cdot\bm{j}'_{H}=j_{H}$ respectively.
In other words, the electrical current as a function of $\theta$ should satisfy
\begin{subequations}
  \begin{gather}
    j_{N,\parallel}(2\pi/3-\theta)=j_{N,\parallel}(\theta)\label{eq:mirror_parallel_app}\\
    j_{N,\perp}(2\pi/3-\theta)=-j_{N,\perp}(\theta)\label{eq:mirror_perp_app}\\
    j_{H}(2\pi/3-\theta)=j_{H}(\theta).\label{eq:mirror_hall_app}
  \end{gather}\label{eq:mirror_app}
\end{subequations}

In the AC case, since the applied AC electrical field satisfies $\tilde{\bm{E}}(t;\theta+\pi)=\tilde{\bm{E}}(t+T/2;\theta)$, there is an additional tempo-spatial symmetry $\bm{j}(t;\theta+\pi)=\bm{j}(t+T/2;\theta)$ which reads
\begin{align}
  \bm{j}^{(n)}(\theta+\pi)=&\frac{1}{T}\int_0^{T}\rd t\; \bm{j}(t;\theta+\pi)\re^{in\omega t}\nonumber\\
  =&\frac{1}{T}\int_0^{T}\rd t\; \bm{j}(t+T/2;\theta)\re^{in\omega t}\nonumber\\
  =&\frac{1}{T}\int_{-T/2}^{T/2}\rd t\; \bm{j}(t;\theta)\re^{in\omega t}(-1)^{n}\nonumber\\
  =&(-1)^n\bm{j}^{(n)}(\theta)
\end{align}
in the frequency domain.
When the electrical field reverses in the second half period ($t\to t+T/2$, $\theta\to \theta-\pi$), the unit vectors in the longitudinal and transverse direction also reverse, i.e., $\hat{r}\to\hat{r}''=-\hat{r}$, $\hat{\theta}\to \hat{\theta}''=-\hat{\theta}$.
So the different current components should satisfy
\begin{subequations}
  \begin{gather}
    j^{(n)}_{N,\parallel}(\theta+\pi)=\hat{r}''\cdot\bm{j}^{(n)}(\theta+\pi)=(-1)^{n+1}j^{(n)}_{N,\parallel}(\theta)\label{eq:tempo-spatial_parallel}\\
    j^{(n)}_{N,\perp}(\theta+\pi)=\hat{\theta}''\cdot\bm{j}^{(n)}(\theta+\pi)=(-1)^{n+1}j^{(n)}_{N,\perp}(\theta)\label{eq:tempo-spatial_perp}\\
    j^{(n)}_{H}(\theta+\pi)=\hat{\theta}''\cdot\bm{j}^{(n)}(\theta+\pi)=(-1)^{n+1}j^{(n)}_{H}(\theta)\label{eq:tempo-spatial_hall}
  \end{gather}\label{eq:tempo-spatial}
\end{subequations}

{
\section{Angular Pattern of the Second Order Resistivity}\label{app:sh_pattern}

In the main text, we calculated the angular pattern of the second order current responses, which is the second order conductivity.
However, the quantities directly measured in the experiments are usually resistivity rather than conductivity.
To better compare with the experimental results, we will study the second order resistivity in this part.
For simplicity, we only focus on the weak field limit where the field driving effect is insignificant and the perturbation theory still works.

For convenience, we will assume the system is polarized in the $\mathcal{Q}_3$ valley.
Since the Hall current $\bm{j}_{H}$ is two orders smaller than the normal current $\bm{j}_{N}$ as illustrated in Fig. \ref{fig:currents}, we will ignore the Hall current responses in the following discussion.
Then up to second order of $E$, the current is written as 
\begin{equation}
  j_{a}=\sum_{b}\sigma^{(1)}_{ab}E_b+\sum_{b,c}\sigma^{(2)}_{abc}E_bE_c,
\end{equation}
where $a,b,c=x,y$ are spatial indices, $\sigma^{(1)}$ and $\sigma^{(2)}$ are the first and second order conductivity tensors respectively.
According to the perturbation theory, these conductivity tensors are calculated as 
\begin{subequations}
  \begin{align}
    \sigma^{(1)}_{ab}=&\left(-\frac{e}{\hbar}\right)^2\tau_i\int\frac{\rd^2 k}{(2\pi)^2}\;n^0_{\bm{k}}\partial_{k_a}\partial_{k_b}\varepsilon^0_{\bm{k}},\\
    \sigma^{(2)}_{abc}=&\left(-\frac{e}{\hbar}\right)^3\tau_i^2\int\frac{\rd^2 k}{(2\pi)^2}\;n^0_{\bm{k}}\partial_{k_a}\partial_{k_b}\partial_{k_c}\varepsilon^0_{\bm{k}},
  \end{align}
\end{subequations}
which are symmetric with respect to the spatial indices, i.e., $\sigma^{(1)}_{ab}=\sigma^{(1)}_{ba}$ and $\sigma^{(2)}_{abc}=\sigma^{(2)}_{acb}=\sigma^{(2)}_{bac}$.
Besides, due to the restriction of the mirror symmetry with respect to the $x$-axis, we have
\begin{subequations}
  \begin{gather}
    j_{x}(E_x,-E_y)=j_{x}(E_x,E_y),\\
    j_{y}(E_y,-E_y)=-j_{y}(E_x,E_y),
  \end{gather}
\end{subequations}
which requires $\sigma^{(1)}_{xy}=0$ and $\sigma^{(2)}_{xxy}=\sigma^{(2)}_{yyy}=0$.
With the only non-zero components of the conductivity tensors, the relation between the current density and electrical field is written as 
\begin{subequations}
  \begin{gather}
    j_{x}=\sigma_{xx}^{(1)}E_x+\sigma_{xxx}^{(2)}E_x^2+\sigma^{(2)}_{xyy}E_y^2\\
    j_{y}=\sigma^{(1)}_{yy}E_y+2\sigma^{(2)}_{xyy}E_xE_y.
  \end{gather}\label{eq:conductivity}
\end{subequations}
Similarly, the electrical field $\bm{E}$ could also be written as a series of $\bm{j}$
\begin{subequations}
  \begin{gather}
    E_x=\rho^{(1)}_{xx}j_x+\rho^{(2)}_{xxx}j_x^2+\rho^{(2)}_{xyy}j_y^2,\\
    E_y=\rho^{(1)}_{yy}j_y+2\rho^{(2)}_{xyy}j_xj_y,
  \end{gather}\label{eq:resistivity}
\end{subequations}
where $\rho^{(1)}$, $\rho^{(2)}$ are the first and second order resistivity tensors.
Substitute Eq. \eqref{eq:resistivity} into Eq. \eqref{eq:conductivity} and keep up to the second order of $j$ we have 
\begin{subequations}
  \begin{align}
    j_x=&\rho^{(1)}_{xx}\sigma^{(1)}_{xx}j_x+\{\rho^{(2)}_{xyy}\sigma^{(1)}_{xx}+[\rho^{(1)}_{yy}]^2\sigma^{(2)}_{xyy}\}j_y^2\nonumber\\
    &+\{\rho^{(2)}_{xxx}\sigma^{(1)}_{xx}+[\rho^{(1)}_{xx}]^2\sigma^{(2)}_{xxx}\}j_x^2,\\
    j_y=&\rho^{(1)}_{yy}\sigma^{(1)}_{yy}j_y+2[\rho^{(1)}_{xx}\rho^{(1)}_{yy}\sigma^{(2)}_{xyy}+\rho^{(2)}_{xyy}\sigma^{(1)}_{yy}]j_xj_y.
  \end{align}\label{eq:resistivity_conductivity}
\end{subequations}
Eq. \eqref{eq:resistivity_conductivity} holds for any $j_x,j_y$, thus the coefficients of the series expansion satisfy the following system of equation
\begin{subequations}
  \begin{gather}
    \rho^{(1)}_{xx}\sigma^{(1)}_{xx}=1,\\
    \rho^{(2)}_{xyy}\sigma^{(1)}_{xx}+[\rho^{(1)}_{yy}]^2\sigma^{(2)}_{xyy}=0,\\
    \rho^{(2)}_{xxx}\sigma^{(1)}_{xx}+[\rho^{(1)}_{xx}]^2\sigma^{(2)}_{xxx}=0,\\
    \rho^{(1)}_{yy}\sigma^{(1)}_{yy}=1,\\
    \rho^{(1)}_{xx}\rho^{(1)}_{yy}\sigma^{(2)}_{xyy}+\rho^{(2)}_{xyy}\sigma^{(1)}_{yy}=0.
  \end{gather}
\end{subequations}
By solving the previous system of equation, the resistivity tensors are expressed by the conductivity tensors as 
\begin{subequations}
  \begin{align}
    \rho^{(1)}_{xx}=&[\sigma^{(1)}_{xx}]^{-1}\\
    \rho^{(1)}_{yy}=&[\sigma^{(1)}_{yy}]^{-1}\\
    \rho^{(2)}_{xxx}=&-\sigma^{(2)}_{xxx}[\sigma^{(1)}_{xx}]^{-3}\\
    \rho^{(2)}_{xyy}=&-\sigma^{(2)}_{xyy}[\sigma^{(1)}_{xx}]^{-1}[\sigma^{(1)}_{yy}]^{-2}
  \end{align}
\end{subequations}

In Ref.\citep{linSpontaneousMomentumPolarization2023}, the angular resolved second order resistivity in the longitudinal direction is measured.
When current $\bm{j}=j(\cos\theta,\sin\theta)$ is applied in the $\theta$ direction, the longitudinal electrical field response is 
\begin{align}
  E_{\parallel}=&E_x\cos\theta+E_y\sin\theta\nonumber\\
  =&j[\rho^{(1)}_{xx}\cos^2\theta+\rho^{(1)}_{yy}\sin^2\theta]\nonumber\\
  &+j^2[\rho^{(2)}_{xxx}\cos^2\theta+3\rho^{(2)}_{xyy}\sin^2\theta]\cos\theta.
\end{align}
Thus the second order longitudinal resistivity as a function of theta is 
\begin{equation}
  \rho^{(2)}_{\parallel}=[\rho^{(2)}_{xxx}\cos^2\theta+3\rho^{(2)}_{xyy}\sin^2\theta]\cos\theta.\label{eq:resistivity_longitudinal}
\end{equation}
Define a dimensionless parameter 
\begin{equation}
  \alpha\equiv \frac{3\rho^{(2)}_{xyy}}{\rho^{(2)}_{xxx}}=\frac{3[\sigma^{(1)}_{xx}]^2\sigma^{(2)}_{xyy}}{[\sigma^{(1)}_{yy}]^2\sigma^{(2)}_{xxx}}.
\end{equation}
Then, according to Eq. \eqref{eq:resistivity_longitudinal}, the angular pattern of the second order longitudinal resistivity $\rho^{(2)}_{\parallel}$ is totally determined by $\alpha$.
At fixed hold density $n_h=0.01\times10^{12}\mathrm{cm}^{-2}$, the parameter $\alpha$ is calculated and plotted as a function of interlayer bias potential $U$ in Fig. \ref{fig:pattern_potential}(a).
In Fig. \ref{fig:pattern_potential}(b-d), the angular patterns of the second order longitudinal resistivity are also plotted at different interlayer potential $U$.
When the interlayer potential $U$ is small and $\alpha<1$, the resistivity takes its maximum value at the polarization direction and the angular pattern shown in Fig. \ref{fig:pattern_potential}(b) agree quantitatively with that in Ref. \citep{linSpontaneousMomentumPolarization2023}(FIG.2.b).
With the increase of the interlayer potential $U$, the parameter $\alpha$ also grows.
When $\alpha>1$, $\rho^{(2)}_{\parallel}$ has two maximum directions distributed symmetrically on the two sides of the mirror line (when the system is polarized in the $\mathcal{Q}_3$ valley, the mirror line is the $x$-axis), and the typical angular pattern is illustrated in Fig. \ref{fig:pattern_potential}(d).
At the critical value $\alpha=1$, the angular pattern is also plotted in Fig. \ref{fig:pattern_potential}(c) for reference.

\begin{figure}[h]
  \centering
  \includegraphics[width=\linewidth]{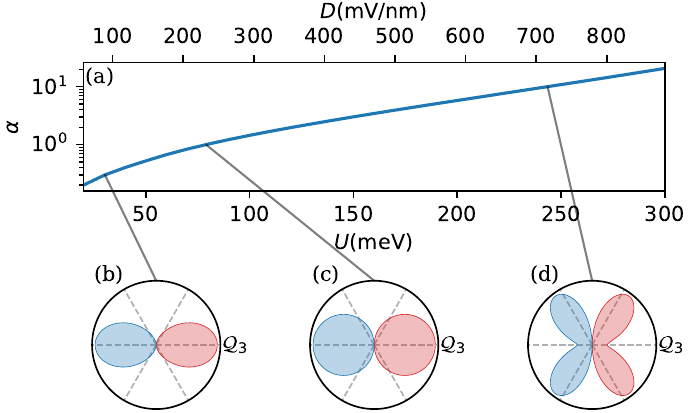}
  \caption{{(a) The dimensionless parameter $\alpha$ as a function of interlayer potential $U$.
  The hole density is fixed at $n_h=0.01\times10^{12}\mathrm{cm}^{-2}$ where the system is in the 1-pocket phase. 
  (b-d) The angular patterns of the second order longitudinal resistivity $\rho^{(2)}_{\parallel}(\theta)$ at different interlayer potential $U$.
  The system is assumed to be polarized in the $\mathcal{Q}_3$ valley.}
  }
  \label{fig:pattern_potential}
\end{figure}

}


%


\end{document}